\documentclass[preprint,12pt]{elsarticle}

\usepackage{amssymb}
\usepackage{amsmath}
\usepackage{bm}
\usepackage{amsthm}
\usepackage{lineno}
\usepackage{url}
\usepackage{xcolor}
\usepackage{caption}
\usepackage{subcaption}
\usepackage{colortbl}
\usepackage{geometry}
\usepackage{graphicx}
\usepackage[acronym,nogroupskip,toc]{glossaries}
\usepackage{nomencl}
\makenomenclature
\usepackage{float}
\usepackage{flafter} 
\usepackage{tabularx}
\usepackage{color,soul}
\usepackage{makecell}

\usepackage[breaklinks=true]{hyperref}

\hypersetup{
    pdftitle={Measurement of Generative AI Workload Power Profiles for Whole-Facility Data Center Infrastructure Planning},
    pdfauthor={Roberto Vercellino}
}

\newcommand{\doi}[1]{{doi:~\href{http://doi.org/#1}{#1}}}

\usepackage{txfonts}

\journal{arXiv}

\begin{document}

\begin{frontmatter}

\title{Measurement of Generative AI Workload Power Profiles for Whole-Facility Data Center Infrastructure Planning}



\author{Roberto Vercellino\corref{cor1}}

\cortext[cor1]{Corresponding author: \href{mailto:roberto.vercellino@nlr.gov}{\texttt{Roberto.Vercellino@nlr.gov}}}
\author{Jared Willard}
\author{Gustavo Campos}
\author{Weslley da Silva Pereira} 
\author{Olivia Hull} 
\author{Matthew Selensky} 
\author{Juliane Mueller} 
\affiliation{
    organization={National Laboratory of the Rockies (NLR)},     
    city={Golden},
    state={CO},
    country={USA}
}


\begin{abstract}

The rapid growth of generative artificial intelligence (AI) has introduced unprecedented computational demands, driving significant increases in the energy footprint of data centers. 
However, existing power consumption data is largely proprietary and reported at varying resolutions, creating challenges for estimating whole-facility energy use and planning infrastructure.
In this work, we present a methodology that bridges this gap by linking high-resolution workload power measurements to whole-facility energy demand. 
Using NLR's high-performance computing data center equipped with NVIDIA H100 GPUs, we measure power consumption of AI workloads at 0.1-second resolution for AI training, fine-tuning and inference jobs. 
Workloads are characterized using MLCommons benchmarks for model training and fine-tuning, and vLLM benchmarks for inference, enabling reproducible and standardized workload profiling. 
The dataset of power consumption profiles is made publicly available.
These power profiles are then scaled to the whole-facility-level using a bottom-up, event-driven, data center energy model. 
The resulting whole-facility energy profiles capture realistic temporal fluctuations driven by AI workloads and user-behavior, and can be used to inform infrastructure planning for grid connection, on-site energy generation, and distributed microgrids.

\end{abstract}


\end{frontmatter}


\begin{table}
\caption*{List of Acronyms}
\begin{tabularx}{\textwidth}{XX}
    \hline
    \textbf{Acronym} & \textbf{Meaning} \\
    \hline
    AI & Artificial Intelligence \\
    CPU & Central Processing Unit \\
    GPU & Graphics Processing Unit \\
    GenAI & Generative Artificial Intelligence \\
    HPC & High-Performance Computing \\
    IT & Information Technology \\
    LLM & Large Language Model \\
    NLR & National Laboratory of the Rockies \\
    TDP & Thermal Design Power \\    
    \hline
\end{tabularx}
\end{table}





\section{Introduction}
\label{sec:introduction}

Data centers have become one of the fastest-growing sources of electricity demand in the United States and globally, driven by the rapid adoption of graphics processing unit (GPU)-accelerated servers for artificial intelligence (AI) and generative AI (GenAI) workloads. 
By 2018 data centers consumed approximately 76~TWh in the U.S., or about 2\% of the total domestic annual electricity consumption~\cite{Shehabi2024}. 
Since then, demand has accelerated markedly, reaching roughly 176~TWh in 2023, accounting for 4.4\% of the U.S. electricity use.
A similar trend is observed globally, with data center demand growing at 12\% per year over the last five years and reaching 415~TWh or 1.5\% of the world’s electricity consumption in 2024~\cite{IEA2025EnergyAI}.
Projections suggest that U.S. data center electricity usage could double or even triple by 2028~\cite{Shehabi2024, DOE2024}, with similar estimates for the rest of the world \cite{IEA2025EnergyAI}.


AI-focused data center deployment comes with its own set of technical, regulatory and socioeconomic challenges, affecting a broad set of entities including technology manufacturers, site developers, utility companies, regulators and communities. 
Many of the challenges are driven by the energy requirements of AI data centers and occur across different operational timescales~\cite{chen_electricity_2025}. 
At the planning level (years to decades) challenges include grid infrastructure upgrades and capacity planning for generation, transmission, and distribution, as well as data center siting, equipment lead times, interconnection permitting and construction processes.
During day-to-day operation (minutes to hours), uncertainty and variability in data center utilization driven by user demand and workload profiles make load forecasting difficult, which in turn complicates generation dispatch and reserve planning, driving up energy costs. 
Real-time operation (milliseconds to minutes) can be hindered by MW-scale load fluctuations resulting from coordinated utilization of thousands of processing devices, mainly GPUs, during large-scale AI training workloads~\cite{choukse_power_2025}.
This variability in power can create grid stability concerns due to voltage fluctuations, harmonics distortions, and resonant events.
Finally, to protect expensive equipment in response to temporary grid disturbances, automatic switches in data centers might shift gigawatts of loads off the grid in a matter of seconds, creating cascading grid instability.
The North American Electric Reliability Corporation (NERC) reported on one such event in 2024 at a data center in Virginia, where a 1.5 GW load drop occurred within 82 seconds, after a 230 kV line fault~\cite{nerc_2025}.

This range of planning, short-term, and real-time energy challenges associated with AI data centers could be mitigated through accurate data center load modeling.
However, developing such models remains inherently difficult for several reasons.
First, large-scale cloud computing providers, also referred to as hyperscalers, rarely disclose facility operation data and loads, as these are considered business-sensitive information.
Medium and small-scale facilities, commonly known as colocation data centers, which lease space and computing resources to multiple commercial user-groups, have often limited visibility into usage and power consumption at the rack- or server-level.
Finally, uncertainty in load modeling is further compounded by the rapidly evolving technology landscape.
On the algorithmic side, AI methods continue to evolve -- some of the latest additions being agentic AI frameworks which are introducing more dynamic resource utilization behaviors~\cite{bodra2025machine}. 
At the infrastructure level, data center architectures are also changing to accommodate higher rack power densities, in part enabled by a transition to high-voltage direct current (HVDC) architectures that improve energy and computational efficiency~\cite{chen_data_2023}.
Together, these shifts not only make load modeling at any scale more difficult, but also risk exacerbating the aforementioned operational and planning challenges.


\subsection{Whole-Facility Data Center Load Modeling}


Prior studies on AI data center load modeling address demand at disparate temporal and system scales, and yet remain  insufficient for facility-level, geographically-resolved power system planning.
Mytton~\cite{MYTTON20222032} reviews 258 estimates of data center energy consumption, finding that methods vary widely in rigor and transparency and result in a wide range of estimates. 
Approaches generally fall into three categories: bottom-up models, top-down aggregation from government statistics, and extrapolations that forecast future demand. 
The review notes that many studies rely on proprietary data, making validation difficult.
Using a similar classification, Shehabi et al.~\cite{Shehabi2024} provide a bottom-up model for estimating data center loads by disaggregating information technology (IT) equipment categories (AI servers, storage, networking) and accounting for non-IT power demands such as cooling and water use. 
The authors use this methodology to estimate load growth in the data center energy sector as a function of time broken down by market segment~\cite{Shehabi2024}. 
While these studies can help estimate overall infrastructure goals, they lack sufficient temporal and geographical resolution to support geographically-resolved generation, transmission and distribution planning.

A literature source from a collaboration of hyperscale compute providers highlights challenges from large-scale large language model (LLM) training at the whole-facility scale~\cite{choukse_power_2025}. However, the power profile is normalized and features only a few minutes in operation, resulting in limited practical use.
Finally, other examples in the literature have captured the power conversion steps between servers and the grid distribution through detailed models of power electronics -- also referred to as electromagnetic transient (EMT) models~\cite{sun_dynamic_2022, ross_electromagnetic_2026}.
These models allow limited flexibility to explore alternative (and future-looking) power distribution architectures, are often computationally expensive and are not informed by AI-specific server power profiles.
Mughes et al. use machine learning to perform short-term load forecasting based on Massachusetts Institute of Technology (MIT)'s publicly available dataset while also demonstrating the complexity of this task~\cite{mughees_short-term_2025}. 

\subsection{Characterization of AI Workloads}

Several experimental studies have measured the instantaneous power draw of GPU-accelerated servers under AI workloads.
In one such study, Latif et al.~\cite{latif2024} measure the power of NVIDIA H100 HGX nodes during image classification (ResNet) and LLM (Llama-2 13B) training. 
They report a peak power of 8.4~kW, 18\% below the manufacturer-rated 10.2~kW, and show that increasing training batch size significantly alters power and energy use.
Such fine-grained measurements are critical to developing accurate bottom-up models that reflect workload-specific variability. However, the dataset is not made publicly available, limiting reproducibility and broader reuse.  

In a complementary work, Patel et al.~\cite{Patel2024} characterize power consumption behavior of LLM workloads, both training and inference, in cloud data center settings. Their measurements span multiple levels of aggregation, including GPUs, server, and rack or row, and reveal distinct phases in inference (e.g., “prompt” vs “token‑generation” phases) and show how power‑management knobs like GPU frequency locking and power capping can reclaim headroom, which in this context defines the gap between the processing unit's power consumption and maximum power rating -- or thermal design point (TDP). Notably, the study finds that training has very little oversubscription headroom ($\sim$3\%) due to large synchronous power peaks, whereas inference, despite high per‑server peaks, have significantly more headroom ($\sim$21\%) and thus are strong candidates for power oversubscription. The authors also introduce the framework POLCA that enables safe power oversubscription in LLM inference clusters, boosting effective capacity by $\sim$30\% with minimal throttling. 

In parallel, several publicly available datasets capture real-world AI workload submission patterns across diverse hardware, deployment settings and customer type, although do not offer detailed insight into their power consumption.
The Azure LLM Inference dataset from Microsoft offers real-world tracking of LLM inference traffic over durations of nine days, which include more than 44 million inference requests \cite{azure_llm_paper, azure_llm_dataset}.
The ACME Trace dataset from the Shanghai AI Lab captures extensive GPU-based workloads for LLM pretraining, inference, and supervised fine-tuning (SFT) over a six-month period from March 2023 to August 2023, totaling 880 thousand job traces \cite{acme_paper, acme_dataset}.
Similarly, the Alibaba GPU Traces dataset records activity across 1,800 machines and 6,500 GPUs over two months \cite{alibaba_paper, alibaba_dataset}. 
In contrast, BurstGPT is a workload trace dataset drawn from a regional Azure OpenAI GPT service, spanning 213 days and more than 10 million requests~\cite{burstgpt_paper, burstgpt_dataset}.
It characterizes real-world LLM-serving patterns, including burstiness in concurrency, conversation session structure, response-length distributions, and failure rates. The paper highlights how using realistic data (as opposed to synthetic workloads) can reveal inefficiencies and guide improvements in scheduling, resource provisioning, and cache management for LLM serving systems. 
A helpful curation of datasets and publications is presented in the Awesome-CloudComputing-Datasets GitHub repository \cite{ACCD} -- although a considerable portion of the listed datasets is outdated (e.g. published before 2020), not GenAI/LLM-related, and does not include the workloads' power consumption.


\subsection{Contributions}
\label{subsec:contributions}

While previous efforts characterizing workload trace of AI submission jobs exist, power consumption is generally not included or underemphasized. 
On the other hand, previous work characterizing power consumption presents  limited transparency and reproducibility, limited combination of algorithm configurations, or does not demonstrate how the profiles could be used to estimate whole-facility data center time-series for infrastructure planning.
This paper bridges these gaps by linking high-resolution power measurements of AI workloads to a whole-facility data center energy model. 
Our contributions are twofold:
\begin{enumerate}
    \item We measure power consumption profiles of GenAI training, fine-tuning and inference workloads at 0.1~s resolution using NVIDIA H100 GPUs. Standardized benchmarks (MLCommons for training and vLLM for inference) ensure reproducibility and representativeness. The dataset is made publicly available at~\cite{osti_3025227}.
    \item We develop a bottom-up, discrete-event, whole-facility data center energy model and demonstrate how to scale-up the measured node-level workload profiles into realistic facility-level computational demand time-series that can be used to perform infrastructure planning.
    
\end{enumerate}


\section{Methodology}
\label{sec:methodology}

\subsection{Overview}

\begin{figure}[ht]
\centering
    \includegraphics[width=\linewidth]{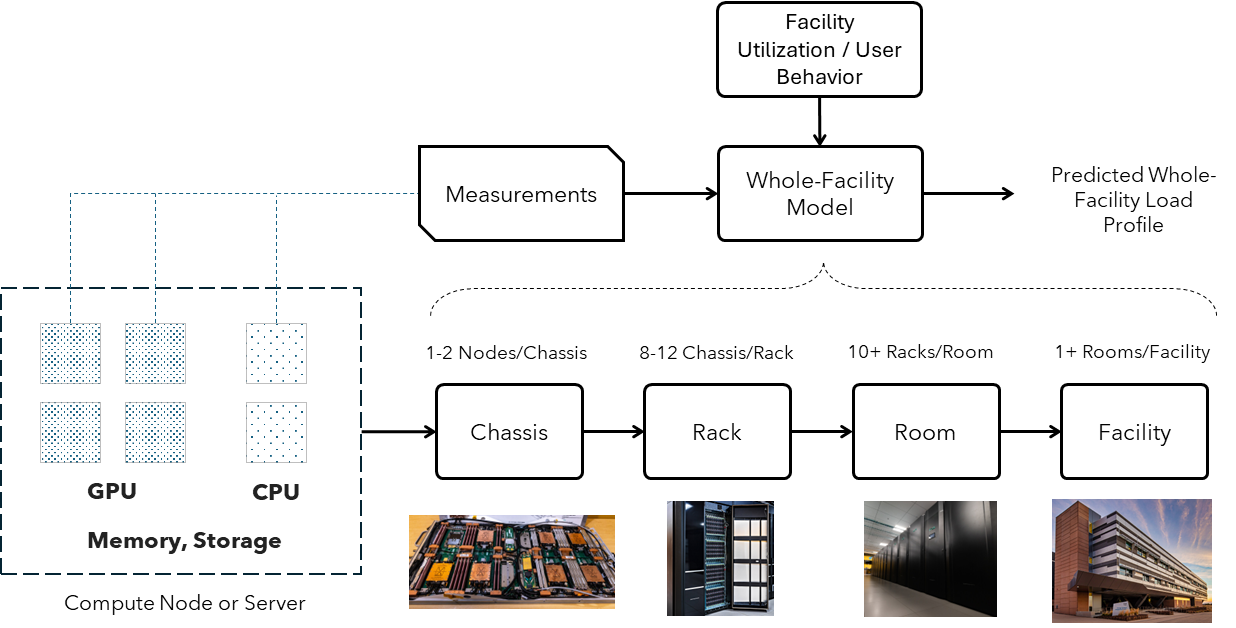}
    \caption{Diagram illustrating the workflow and scaling from data center compute node to facility level\protect\footnotemark.
    }
    \label{fig:node_scale_up}
    \vspace{-15pt}
\end{figure}

\footnotetext{Photos by Agata Bogucka (NLR 104051), Werner Slocum (NLR 63014, 67310), and Joe DelNero (NLR 101334).}

Figure \ref{fig:node_scale_up} illustrates the proposed workflow, linking node-level CPU and GPU power measurements to facility-level simulation, while also capturing the hierarchical levels in data center computational equipment.
Each level is further described next, with corresponding examples from the high-performance computing (HPC) system at the National Laboratory of the Rockies (NLR) given in parentheses.
A GPU node is composed of CPUs (e.g. 2 AMD EPYC “Genoa” processors with 64 cores per socket), GPUs (e.g. 4 NVIDIA H100 GPUs with 80 GB of VRAM each), RAM system memory (e.g. 384 GB, 768 GB, or 1.5 TB), and local storage (e.g. NVMe ranging from 3.4 TB to 14 TB depending on the RAM tier), all interconnected via PCIe/NVLink communication.
These components define the primary sources of dynamic and workload-dependent power consumption and heat generation at the node-level.

Node or server hardware is physically housed inside of a chassis (e.g. HPE Cray XD), which provides mechanical support, local power distribution, and node-level cooling interfaces.
Typical high-power GPU-accelerated chassis host 1 node per chassis, sometimes 2 in smaller configurations.
The node hardware may be physically adjacent (i.e. on the same board) or distributed throughout the data center and linked via interconnect equipment.
Each chassis is mounted inside of a rack, a standard 42 U rack typically holding $\sim$8–12 chassis -- in this context U is a standard rack unit for IT equipment size and is equivalent to 1.75 inches.
Some space in the racks is reserved for power supply, distribution and cooling equipment, which can be distributed inside of each rack or aggregated in a separate location.
Multiple racks are housed in a single hall or room with controlled temperature, humidity, and dew point for IT server inlets -- ASHRAE TC 9.9 presents thermal guidelines for controlling this space~\cite{ASHRAE2021}.
The same room also houses racks for other purposes such as CPU nodes.

At the facility-level, the aggregation of IT load from all rooms, which can be as large as 90\% of the total site load, is supported by centralized electrical and thermal infrastructure.
Electrical systems typically include medium-voltage utility connections, transformers, uninterruptible power supplies (UPS), switchgear, and backup generators, while cooling systems include centralized equipment such as chillers, cooling towers, pumps, heat exchangers, and piping networks.
Advanced HPC and AI facilities typically employ hybrid air–liquid or fully liquid cooling systems with centralized heat rejection.

\subsection{Measuring Power Profiles of Individual AI Jobs}
\label{subsec:}

\subsubsection{CPU and GPU power measurement}

WattAMeter is an open-source package developed to monitor and record power consumption of CPUs and GPUs over time in a multi-node HPC environment~\cite{wattameter}. 
At its core, WattAMeter provides a “power tracker” abstraction that periodically samples the instantaneous power draw of system components and logs it into time-series data.
Users can use it via a Python API or via a command-line interface, and it supports integration with batch scheduling systems such as Slurm to capture power usage associated with each compute node used by a job (see WattAMeter in~\cite{nlr-hpc}).

Key features include configurable sampling intervals for reading power, frequency for writing the outputs to disk, and options to track metadata like machine ID and job number. For researchers focused on workload profiling (especially AI, HPC), WattAMeter offers a way to collect fine‐granularity power trace data, enabling improved modeling of energy consumption over time, understanding transient behaviors (e.g. GPU power ramp up/down), and validating bottom-up energy models.

WattAMeter is built on top of tools such as NVIDIA Management Library (NVML) and Intel Running Average Power Limit (RAPL). 
These low-level software libraries read or estimate information, such as power, from hardware and have been recurrently used by other researchers for power time-series analysis. 
WattAMeter provides benchmarks to help tune the frequency of update of each low-level reader, as this depends on the hardware.

We found a few sources of evidence to support the accuracy of those tools in the literature. 
NVIDIA documentation~\cite{nvml-docs} suggests an error of ±5\% in NVML power draw, which is verified in~\cite{10.1109/SC41406.2024.00028}. 
In~\cite{10.1145/3177754}, the authors verify that RAPL has a correlation of 99\% with AC power measurements in Intel CPUs, although no similar sources were found for AMD CPUs. 
Intel RAPL’s accuracy is usually high in modern CPUs because the values produced by the API are based on hardware measurements, rather than estimates~\cite{10.1145/2989081.2989088}.

\subsubsection{Training and Fine-tuning Benchmarks}

The MLCommons (MLPerf) benchmarks~\cite{mlperf2019} are a widely adopted standard for evaluating AI model training and fine-tuning scenarios. 
Fine-tuning benchmarks specifically measure the computational and power requirements of adapting pre-trained models to new datasets or tasks, rather than training from scratch. 
These benchmarks include diverse model and problem types, such as Large Language Models (LLM) (e.g. Llama-2 70B, GPT3), Natural language processing (NLP) (e.g. Llama-3.1 8B, BERT-large), image classification and computer vision. 

In our study, we focused on two different AI models both from the MLPerf Training v4.0 suite~\cite{mlperf-training-v4.0}.
First, we evaluated the fine-tuning of the Llama-2 70B parameter LLM model developed by Meta~\cite{touvron2023Llama2}, which is representative of large-scale transformer workloads. 
The Llama-2 70B model uses a transformer architecture optimized for generative natural language tasks. 
The benchmark uses the SCROLLS GovReport dataset~\cite{shaham2022scrolls} and a PyTorch implementation framework. 
Fine-tuning this model involves using Low-Rank Adaptation (LoRA)~\cite{hu2022lora}, a parameter-efficient technique that updates low-rank matrices in the attention and feedforward layers while keeping the original model weights frozen. 
This approach significantly reduces the memory and computational cost while maintaining performance comparable to full fine-tuning. 
It also leverages DeepSpeed ZeRO-3 parameter offloading~\cite{ren2021zero}, which partitions all model states (parameters, gradients, and optimizer states) across ranks and offloads them to CPU memory to drastically lower per-GPU memory usage.

In the MLCommons reference implementation used in this study, distributed scaling was achieved through data parallelism combined with ZeRO-3 state partitioning, rather than tensor or pipeline parallelism. 
The benchmark was configured with both per-device training batch size equal to 1 as well as a gradient accumulation factor of 1, such that the effective global batch size scaled linearly with the number of GPUs used. 
In our experiments, the compute resources were scaled from 2 to 4, 8, and 16 nodes, resulting in a corresponding increase in the global batch size at each scale.
This represented weak-scaling with respect to the optimization step, where the work each GPU was doing remained the same while the aggregate work per optimization step increased with the number of nodes. 
Although ZeRO-3 reduces per-GPU memory requirements, it does not alter the underlying data-parallel training. 
As a result, larger-scale runs will have higher communication volume compared to smaller configurations due to the increase in the number of data-parallel ranks.

We note that while a higher per-device batch size could be used and still fit in device memory, especially for the higher node counts where ZeRO-3 partitioning effectively reduces per-device model state memory, we kept the value at 1 to match the default in the MLPerf implementation. We found that higher batch sizes introduced instability in the training loss curve that was not easily corrected by reducing the learning rate proportionally. With a proper hyperparameter sweep over the learning rate and batch size, this would be expected to increase multi-node efficiency and decrease computational time, but that was outside the scope of this study.

A second MLPerf benchmark, the Stable Diffusion v2, was also evaluated. 
This benchmark trains a generative image model to a fixed quality target using a large text-to-image dataset. 
It uses a PyTorch implementation framework and the LAION-400M-filtered dataset~\cite{schuhmann2021laion}, with roughly 865 million parameters in the model.
The workload consists of UNet-based denoising training with variational autoencoder and text encoder components executed in each iteration, producing a mixture of convolutional, attention, and encoder-decoder operations that differs substantially from the transformer-only Llama-2 workload. 
For this reason, this Stable Diffusion v2 makes a valid complementary representative workload for AI datacenter modeling. 
The MLCommons reference implementation of Stable Diffusion v2 also relies on data-parallel training to scale across multiple GPUs and nodes. 
In our experiments, the workload was scaled from 1 to 2, 4, 8, and 16 nodes, resulting in an increase in the global batch size as the number of data-parallel ranks increased, also indicating weak scaling. 
More details on the training benchmark architectures can be found in~\cite{mlcommons_training_github}.

\subsubsection{Inference Benchmarks}\label{sec:met-inference}

For inference profiling, we relied on vLLM~\cite{vllmkwon2023efficient}, an optimized inference engine designed for LLMs. vLLM implements a highly efficient PagedAttention mechanism~\cite{vllmkwon2023efficient}, which reduces memory fragmentation and enables large batch sizes without compromising latency.

Inference models are highly dependent on the number of input and output tokens, defined as numeric representations of text fragments. 
For instance, a heuristic rule of 1.33 tokens per word (equivalently, 100 tokens per 75 words) applies to the English language~\cite{openai_tokens}. 
The model limit is usually referred to as the context length, defined as the sum of input and output tokens. 
Typical context length values include 4,096 tokens for OpenAI's GPT-3.5 (later updated to 8,192) and 8,192 tokens for GPT-4, 128,000 tokens for GPT-4o and GPT-4o mini, 4,096 tokens for Meta's Llama-2~\cite{IBMContextWindow2025}.
Examples of maximum output-token limits include 4,096 for GPT-4 and GPT-4-Turbo, and 16,384 tokens for GPT-4o and GPT-4o mini~\cite{IBMContextWindow2025}.
However, lower input and output token constraints are usually enforced by LLM cloud providers; they might be adjusted based on the license tier, or even dynamically during a session.

Inference benchmarks with Llama-3 70B focus on (1) token throughput (tokens generated per second per GPU), which determines system-level capacity for serving user queries, (2) end-to-end latency across varied batch sizes and sequence lengths, reflecting real-time application requirements, (3) memory efficiency, where vLLM’s caching and tensor reuse mechanisms reduce overhead compared to naive inference engines, and (4) scaling across GPUs, allowing evaluation of multi-node performance in HPC or data center environments.

From a power perspective, inference workloads produce a different profile compared to fine-tuning, characterized by shorter bursts of GPU activity during forward passes, often interspersed with idle or low-utilization periods when waiting for input or handling smaller batches~\cite{li_ai_2025}. Capturing these dynamics complements the training power traces and allowed us to construct whole-facility load models that combine diverse job mixes.

Another key distinction for inference benchmarks is between offline (batch) vs. online (serve) modes~\cite{vllm_quickstart_2025}. 
The offline mode processes a fixed or static batch of requests that are passed to the model at the same time, while the online mode processes dynamic, user-driven requests that arrive at different times and are batched using a dynamic or adaptive strategy. 
The former maximizes tokens per second (TPS) or throughput by optimizing token generation order for maximum GPU utilization, typically used in benchmarks to characterize peak system performance, while the latter minimizes request latency for client satisfaction. 
In terms of power draw, the offline mode is expected to consume close to the device's Thermal Design Power (TDP) for shorter periods of time and have higher efficiency (lower energy per token), while the online mode is expected to have larger variations in time but a smaller average consumption and lower efficiency (higher energy per token). 
In order to estimate realistic power draw for production systems, the online (serve) mode is preferred.








\subsection{Whole-Facility Simulation - DIPLOEE}\label{sec:model_structure}

The Data Center Infrastructure Planning, Load Optimization for Energy Efficiency (DIPLOEE) model combines server-level data center workload power samples with facility utilization distributions to simulate data center operation across the whole facility.
The model takes a slightly different structure depending on the workload type being simulated.
First, we describe a job-centered architecture, useful for representing training and fine-tuning jobs typical of AI model development and colocation data centers.
Then we present an architecture better suited to simulate operation for clusters hosting inference models in production.

\subsubsection{Job-Centered Simulation}\label{sec:DIPLOEE_job}

\begin{figure}[ht]
\centering
    \includegraphics[width=\linewidth]{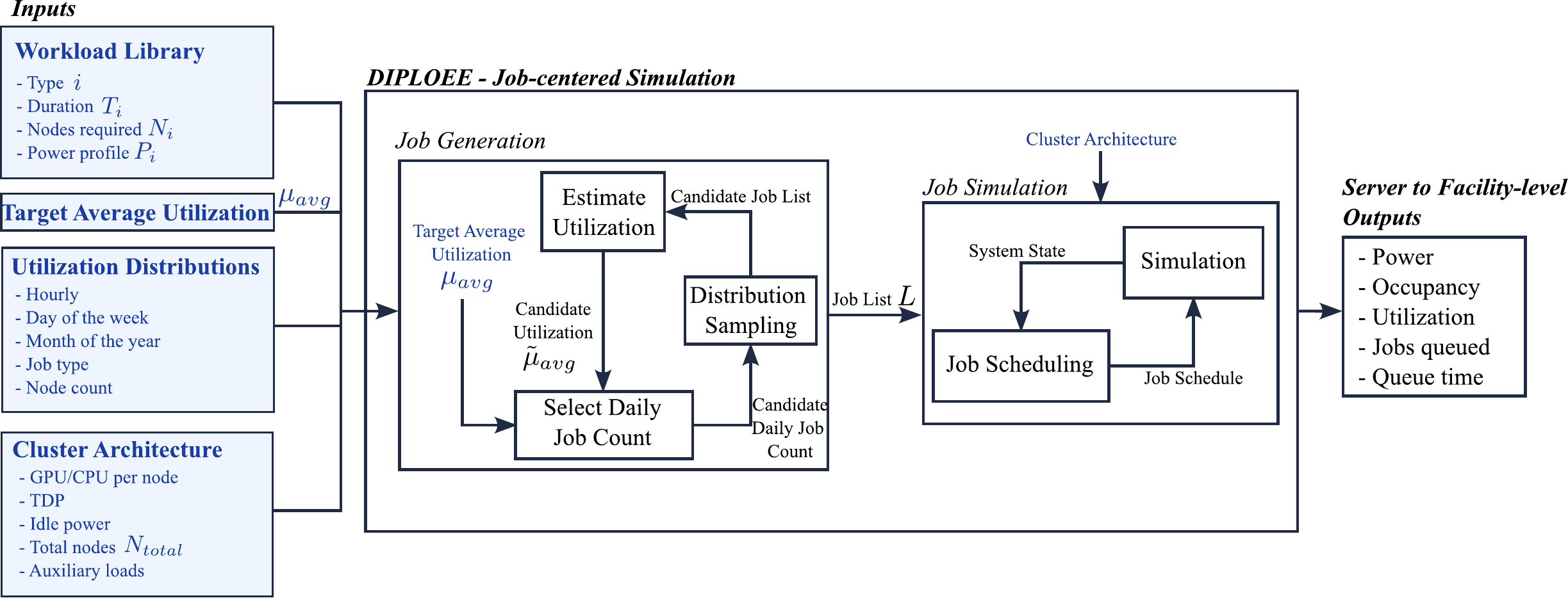}
    \caption{Data Center Infrastructure Planning, Load Optimization for Energy Efficiency (DIPLOEE) model diagram.}
    \label{fig:diploee_diagram-colocation}
\end{figure}

The diagram in Figure~\ref{fig:diploee_diagram-colocation} portrays the main components of DIPLOEE, as well as the flow of data between them. 
The left side of the diagram lists the model's main inputs.
Each workload that might be executed at the data center ($i$) is uniquely defined by type, number of nodes used ($N_i$), duration ($T_i$), and most importantly, time-resolved power consumption profile ($P_i$).
To determine facility usage -- i.e. which workloads are submitted and when -- the model uses probability distributions, including workload type and node count, as well as temporal distributions such as utilization rate at each time of the day, day of week, and month of the year.

The data center architecture is defined by number of GPUs/CPUs per node, GPU/CPU TDP and number of nodes in the data center.
Another key input is data center target average utilization ($\mu _{avg}$), given by the data center node occupancy as a function of time ($N(t)$), divided by the number of total nodes ($N_{total}$), averaged over the simulation interval ($t_0$, $t_f$).
Mathematically, this can be described as:

\begin{align}\label{eq:utilization}
    \mu _{avg} = \frac{\Delta t}{t_f - t_0}\sum _{t=t_0} ^{t_f} \frac{N(t)}{N_{total}}
\end{align}

It is important to further differentiate the meaning of utilization as defined in Equation~\ref{eq:utilization} from other definitions.
In the context of DIPLOEE, utilization is a normalized measure of the number of nodes currently executing workloads; it does not take into account the efficiency, or the power consumption, of CPUs and GPUs relative to their TDP.

The model uses the specified set of inputs to generate the list of jobs to be simulated in the \textit{Job Generation} step.
This is an iterative process where a bisection algorithm is used to match the user-defined target utilization with an average daily job count.
Each iteration in the search begins with a candidate average daily jobs count ($\tilde{\mu} _{avg}$).
Daily jobs are distributed by day, hour, job type, and node count, based on the user-selected probabilities.
Without fully simulating operation, we estimate average utilization based on the list of jobs and their expected duration.
The error between this estimate and the target utilization is used to refine the average daily job count in the next iteration.

Once the list of jobs ($L$) is generated, this is passed to a discrete event simulation platform built with the Python package SimPy, to simulate the scheduling, execution, and power consumption of workloads on data center nodes, which are represented as limited-resource objects~\cite{scherfke_simpy_2023}.
This step iterates between scheduling and simulation.
The scheduling strategy calculates each workload's scheduled start time: it might consider internal factors such as workload queue time and duration, current and forecasted queue depth and user activity; it might also consider external signals such as dynamic energy pricing, onsite generation or utility demand response signals. 
For this work, we adopted a simple first-in-first-out scheduling strategy, which schedules jobs to run as soon as there are enough resources available and in the order in which the jobs arrived.
When the jobs incur their scheduled start time in the simulation, they are allocated to data center nodes throughout their duration, while data center-wide node occupancy, utilization and power consumption are tracked.
At the end of the simulation, a postprocessing step aggregates node- and job-level series into higher-level metrics, such as (min, max, average, etc.) power consumption, occupancy, utilization and job queuing.

\subsubsection{Inference Simulation}\label{sec:DIPLOEE_inference}

\begin{figure}[ht]
\centering
    \includegraphics[width=\linewidth]{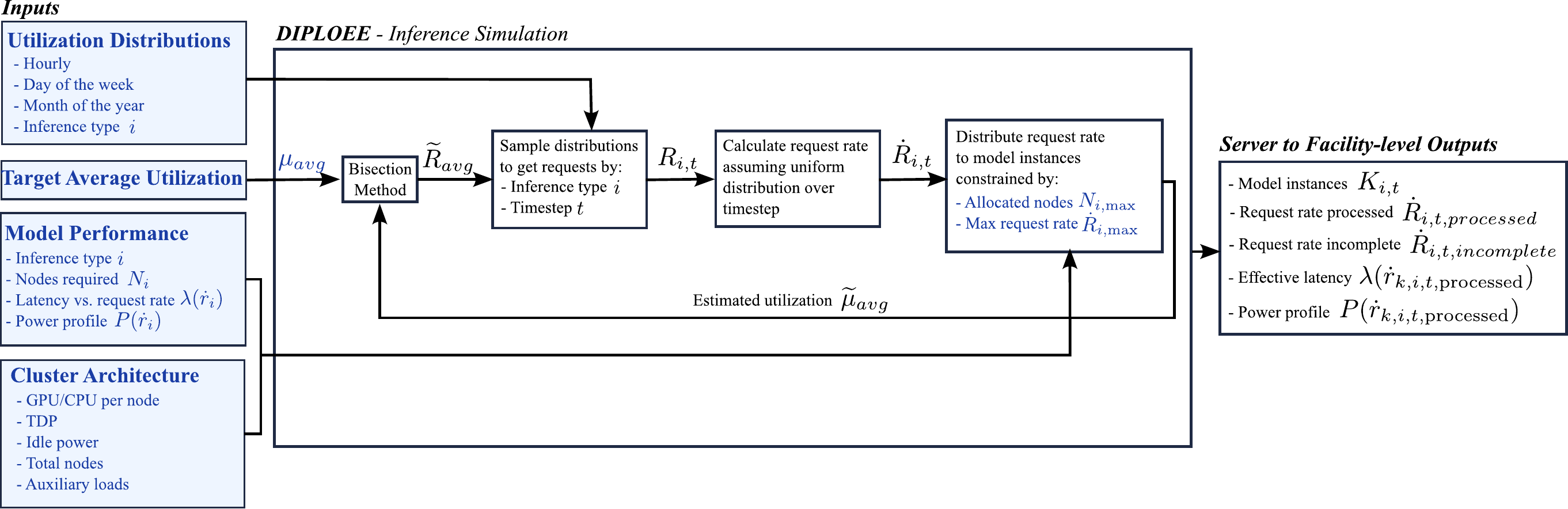}
    \caption{DIPLOEE model details for inference data center use case.}
    \label{fig:diploee_diagram-inference}
\end{figure}

The process outlined in Section~\ref{sec:DIPLOEE_job} was modified to represent inference data centers.
In this case, it is not meaningful to define operation as a function of discrete jobs submitted, scheduled and executed.
Instead, operation is defined by AI models running in online or serve mode, waiting for prompts to be submitted by users who expect limited latency.
Prompts related to different services -- e.g. coding support, web-searches, fraud-detection -- are processed by different inference models which have been specifically trained and fine-tuned for those tasks.
This operating scheme is reflected in Figure~\ref{fig:diploee_diagram-inference}.
In the  DIPLOEE's \textit{Inference Simulation} step, DIPLOEE matches the target utilization $\mu _{avg}$ with an average daily count of inference requests $R_{avg}$. 
Each iteration distributes a candidate number of daily requests $\widetilde{R}_{avg}$ to each timestep ($t$) and among service/inference types ($i$) based on user-provided distributions, such  that:
\begin{align}
    \widetilde{R}_{avg} = \sum_{t}\sum_{i}R_{i,t}
\end{align}
Then, $R_{i,t}$ is converted to an average request rate $\dot{R}_{i,t}$ assuming the requests are uniformly distributed across each timestep. 
Mathematically, this is equivalent to:
\begin{align}
    \dot{R}_{i,t} = \frac{R_{i,t}}{\Delta t}
\end{align}
In general, latency increases with request rate, impacting performance and customer experience.
To limit latency ($\lambda$) the single stream of requests to each inference type is distributed among different model instances served on parallel nodes.
For each inference type, the user defines a maximum request rate per model instance ($\dot{R}_{i,\max}$) which is used to calculate the number of model instances required to satisfy the request rate at each timestep while limiting latency.
The number of model instances is also bounded by a user-defined node count per inference type ($N_{i,\max}$). 
If the total request rate exceeds the maximum number that can be satisfied across nodes, DIPLOEE will track either the increase in latency or record the number of requests not supported, with the latter option selected for this work.
The number of model instances for inference type ($K_{i,t}$) can be expressed as:
\begin{align}
    K_{i,t} = \min \left(\frac{\dot{R}_{i,t}}{\dot{R}_{i,\max}}, N_{i,\max} \right)
\end{align}
where $N_{i,\max}$ is the number of nodes in the cluster reserved for each inference type, and $\dot{R}_{i, \max}$ is the request rate corresponding to the maximum latency allowed:
\begin{align}\label{eq:l_max}
    \dot{R}_{i, \max} = \dot{R}_{i}(\lambda _{i, \max})
\end{align}
We assume the request rate is distributed uniformly across each model instance ($k$), and we indicate the request rate for successfully processed prompts as $\dot{r}_{k, i, t, \mathrm{processed}}$, such that:
\begin{align}
    \dot{R}_{i,t} = \dot{R}_{i, t, \mathrm{incomplete}} + \sum_{k}^{K_{i,t}} \dot{r}_{k, i, t, \mathrm{processed}}
\end{align}
where $\dot{R}_{i,t,\mathrm{incomplete}}$ is the portion of the requests above the maximum rate threshold, which we assume cannot be completed without compromising latency requirements.
While the node count for each inference type can be freely defined by the user, it can be helpful to assign a value proportional to the fraction of requests expected. 
For instance, the total number of nodes ($N_{total}$) might distributed to each inference type ($N_{i, \max}$) using:
\begin{align}\label{eq:inference_nodes_allocation}
    N_{i, \max} = N_{total}
    \frac{p_i  n_i  }{\dot{R}_{i,\max}} 
    \displaystyle\sum_{i}\dfrac{\dot{R}_{i,\max}}{ p_i  n_i}
\end{align}
where $p_i$ is the probability a request falls into a particular inference type and $N_i$ is the nodes required by each model instance for that inference.

Once the target utilization is matched with appropriate synthetic usage data, defining model instances, node occupancy and expected request rate at each timestep ($\dot{r}_{k, i, t, \mathrm{processed}}$), the corresponding load consumption samples ($P(\dot{r}_{i})$) are assigned to each node for each timestep of the simulation.
In contrast with job-oriented simulation, we did not implement any job scheduling under the assumption that minimum latency has ultimate priority.

\section{Results}
\label{sec:results}


\subsection{Experimental Setting}

Experiments were performed on NLR's Kestrel HPC system, which hosts 2,314 CPU-only nodes and 156 GPU-accelerated nodes. 
All experiments were run on the GPU nodes, each housing four NVIDIA H100 SXM GPU accelerators. 
Within a node, each GPU device includes a 900 GB/s NVLink interconnect (18 bidirectional links at 50GB/s) for direct intra-node GPU communication. 
Each node also features two network interface cards (NICs) for communication with other GPU nodes on the HPC system. 
High-speed, low-latency communication between nodes on Kestrel occurs over the HPE Slingshot-11 interconnect, which offers a fabric with peak measured bidirectional bandwidth of 50 GB/s between two nodes, enabling efficient communication between GPU nodes.
Additionally, a given GPU node contains two AMD EPYC 9554 (Genoa) CPUs, each featuring 64 cores, totaling 128 CPU cores per GPU node. 

The GPU nodes offer varying memory and storage configurations to accommodate different workload requirements, with 108 nodes having 384 GB of DDR5 system memory and 3.4 TB of NVMe local storage, 24 nodes having 768 GB of system memory and 3.4 TB of NVMe local storage, and 24 nodes having 1.5 TB of system memory and 14 TB of NVMe local storage. 
Regardless of the volume of system memory on a node, each H100 GPU is configured with 80GB of HBM3 device memory, facilitating the handling of large-scale AI and machine learning tasks. 
The combination of AMD CPUs and NVIDIA GPUs in these nodes allows for high-throughput processing and efficient parallel computation.

The technical specifications for the NVIDIA H100 SXM GPU model can be found in~\cite{nvidiah100}, while the specifications for the AMD EPYC Genoa processors can be found in~\cite{amdepicgenoa}. 
The GPU has a maximum TDP of 700 W, while the CPU has a max TDP of 360 W (and 320--380 W configurable range), meaning that at full capacity (e.g. under a stress test) each node should consume up to 3,520 W (2,800 W from GPUs and 720 W from the CPU, excluding peripherals).
Idle and stress tests for GPUs and CPUs are presented in the~\ref{appA}. 
Results verified that each CPU socket operated at 338.6 W on average when stressed with large dense matrix multiplication kernels. 
The HPL benchmark was used to stress-test GPUs, yielding a mean peak power of 696 W.
Finally, the measured average and standard deviation for idle power of operation were 72.5 W ($\pm$ 0.1 W) for each GPU, and 64.1 W ($\pm$ 4.8 W) for each CPU socket.



\subsection{Llama-2 70B (LLM) Fine-Tuning Workload Power Profiles}\label{sec:exp_finetuning_LL}

\begin{figure}[ht]
\centering
    \includegraphics[width=\linewidth]{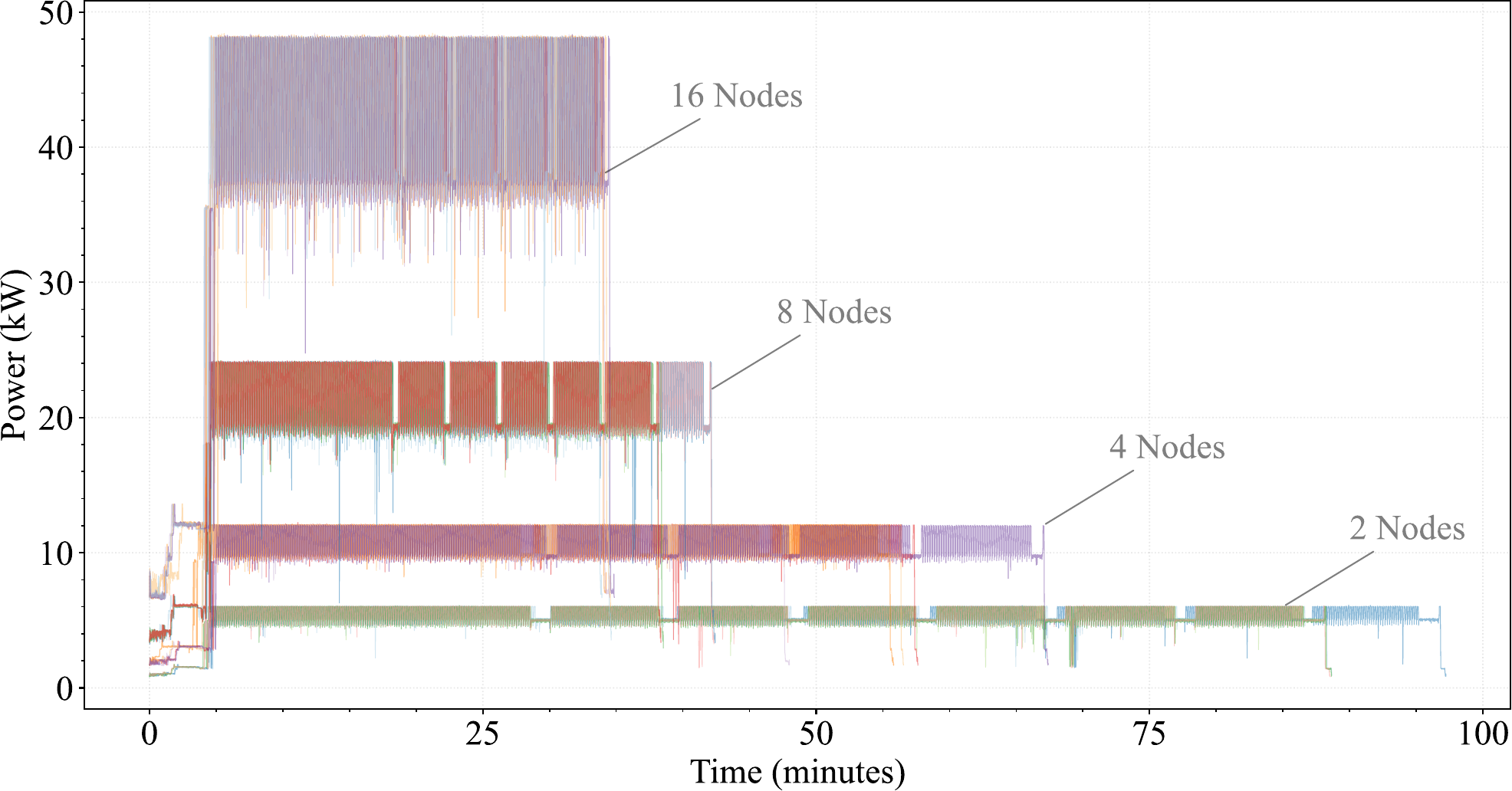}
    \caption{Llama-2 70B fine-tuning power profiles. Replicate experiments using the same node count are shown as superimposed transparent curves with different colors. Each power profile aggregates power consumption across all CPU and GPU devices utilized by the workload.}
    \label{fig:llama2_70b_lora_training_power_profile}
\end{figure}

First, results for the Llama-2 70B fine-tuning benchmark workloads from MLCommons/MLPerf are presented.
Time-resolved power consumption profiles using different numbers of nodes are shown in Figure~\ref{fig:llama2_70b_lora_training_power_profile}.
Across all configurations, an initial ramp-up period of $\sim$3 minutes was observed, corresponding to model checkpoint loading, dataset initialization, CUDA context creation, and ZeRO-3 state partitioning, followed by the fine-tuning regime characterized by high-frequency variations with intermittent low-activity periods in between.
During fine-tuning, the observed power fluctuations were consistent with step-synchronous data-parallel training, in which computation phases on the GPUs alternate with collective communication and CPU to GPU data transfers associated with ZeRO-3 optimizer and state offloading.
These effects were present at all scales and became more pronounced as the node count increased, reflecting the larger number of data-parallel ranks, increased communication volume and synchronization processes at scale.
The profiles highlight transient power dynamics that may be important for facility-level power provisioning and real-time load management.

\begin{figure}[ht]
    \begin{subfigure}{0.49\textwidth}
        \centering
        \includegraphics[width=\linewidth]{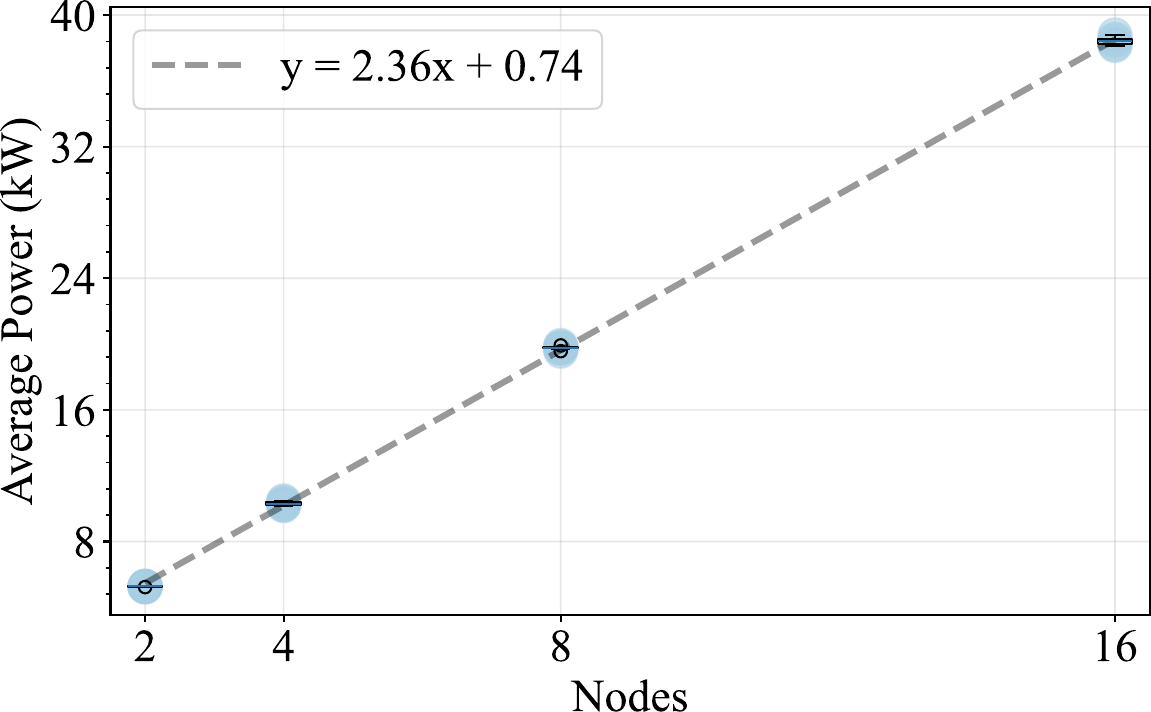}
        \caption{}
        \label{fig:llama2_70b_lora_subplots_power_avg}
    \end{subfigure}
    \begin{subfigure}{0.50\textwidth}
        \centering
        \includegraphics[width=\linewidth]{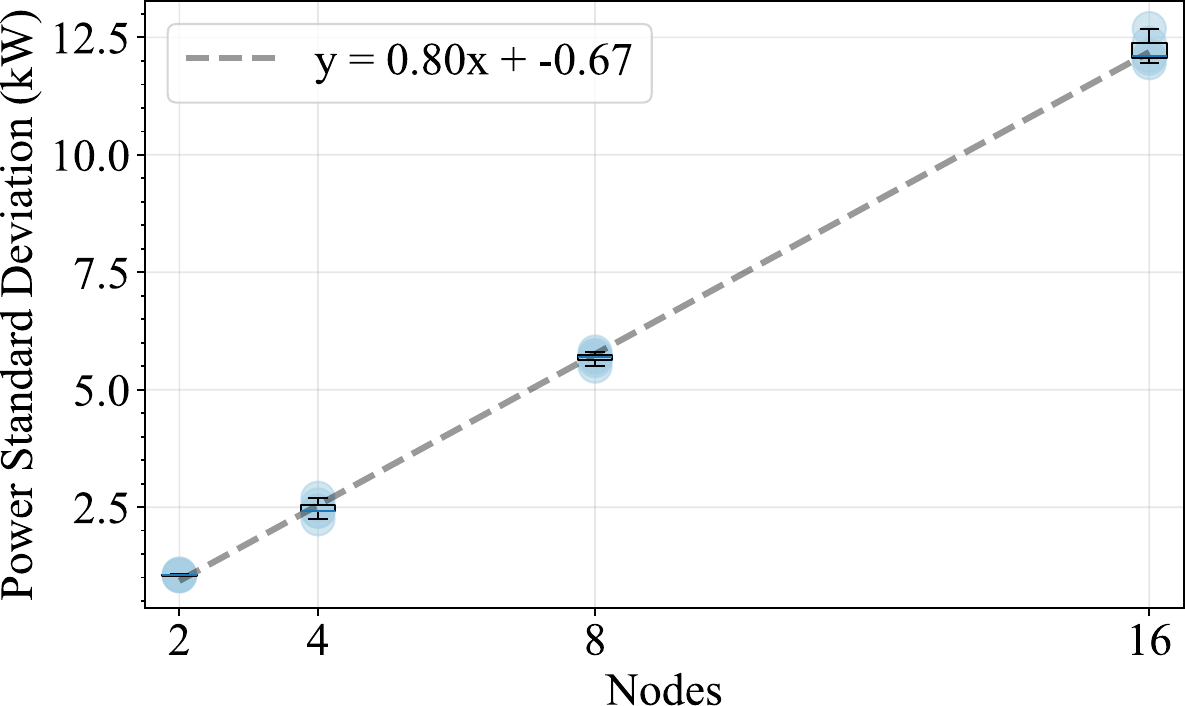}
        \caption{}
        \label{fig:llama2_70b_lora_subplots_power_std}
    \end{subfigure}
    \begin{subfigure}{0.495\textwidth}
        \centering
        \includegraphics[width=\linewidth]{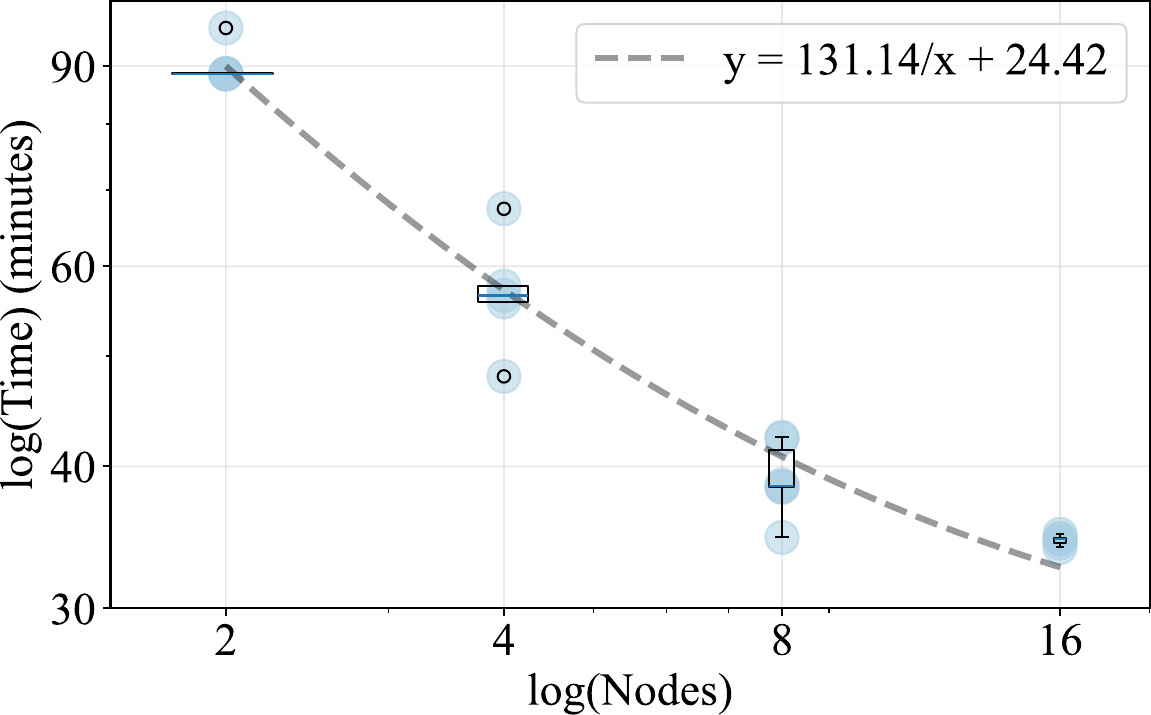}
        \caption{}
        \label{fig:llama2_70b_lora_subplots_time}
    \end{subfigure}
    \begin{subfigure}{0.50\textwidth}
        \centering
        \includegraphics[width=\linewidth]{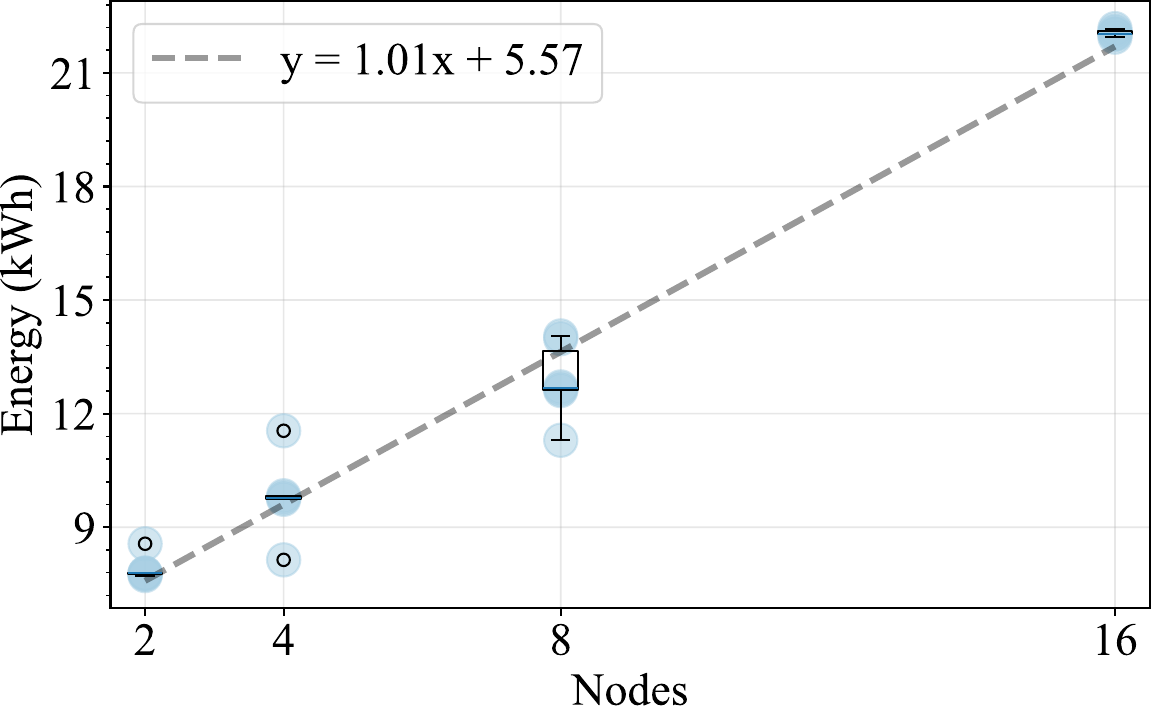}
        \caption{}
        \label{fig:llama2_70b_lora_subplots_energy}
    \end{subfigure}
    \caption{Llama-2 70B fine-tuning energy metrics as a function of number of nodes. Replicate experiments are shown as red dots, vertically distributed with a box-and-whisker plot for each node value. Fitted curves are also presented as dashed lines.}
    \label{fig:llama2_70b_lora_subplots}
\end{figure}

The relationship between node count and power-related metrics for the LLM fine-tuning workload is presented in Figure~\ref{fig:llama2_70b_lora_subplots}.
Subplot~\ref{fig:llama2_70b_lora_subplots_power_avg} shows the linear scaling of the average power consumption with node count, including the initial ramp-up period in the average calculation.
The power variability is illustrated in Subplot~\ref{fig:llama2_70b_lora_subplots_power_std}, characterized by the standard deviation throughout the training session, also exhibiting a linear relationship with the number of nodes.
The wall-clock computational time with the number of nodes is shown in Subplot~\ref{fig:llama2_70b_lora_subplots_time} in log-log scale.
Variance across replicate experiments was noticeably higher than for power consumption.
Subplot~\ref{fig:llama2_70b_lora_subplots_energy} shows the total energy consumption of the fine-tuning sessions, which can be calculated by multiplying the average power in kW by the computational time in hours (equivalent to integrating the profile).

\begin{table}[ht]
    \centering
    \caption{Summary of runtimes from six replicate Llama-2 70B fine-tuning workloads.}
        \begin{tabular}{lrrrrr}
        \label{tab:L2TrainingTimePerEpoch}
        \textbf{Nodes} & \textbf{\thead{Mean Total\\Runtime (sec)}}  & \textbf{\thead{Mean Epoch\\Runtime (sec)}} & \textbf{\thead{Mean Epoch\\Speedup}} & \textbf{\thead{Number of\\Epochs}} & \textbf{\thead{Number of\\Evaluations}} \\
        \hline
        2 & 5,152 ± 226 & 3,930 ± 61 & 0.00 & 0.91 ± 0.04 & 7.20 ± 0.45 \\
        4 & 3,171 ± 403 & 1,353 ± 178 & 2.90 & 1.18 ± 0.14 & 4.00 ± 0.71 \\
        8 & 2,090 ± 170 & 795 ± 50 & 1.70 & 1.51 ± 0.12 & 6.17 ± 0.75 \\
        16 & 1,798 ± 15 & 342 ± 3 & 2.32 & 2.64 ± 0.00 & 5.00 ± 0.00 \\
        \hline
        \end{tabular}
\end{table}

The runtimes shown in Figure~~\ref{fig:llama2_70b_lora_subplots_time} were largely influenced by three factors: the hardware scaling efficiency with respect to the node count, the number of epochs in a given run, and the number of evaluations in a given run, which are presented in Table~\ref{tab:L2TrainingTimePerEpoch}. 
The variance in the number of epochs for different node counts was due to the dependence of the weight-updating optimizer steps on the global batch size. 
Fixing the per-device batch size meant the global batch size increased with the node count, reducing the magnitude of the gradients, which were averaged over more samples in the global batch, thus requiring more epochs and a longer runtime to converge.
The number of evaluations that occur within a training run is a function of both the node count (because the number of evaluations is partially a function of the number of epochs) and the fact that evaluation occurred every 48 training steps, the default in the MLPerf implementation. As larger node counts utilized a larger global batch size, more samples were processed for a given training step, so fewer evaluations were performed overall. Thus, the relative runtime contribution from evaluation decreased as node count increased. 
Finally, the workload scaling in terms of hardware utilization is a function of any efficiency losses due to bottlenecks that generally grow worse as node count increases (e.g., network communication overhead) and the fact that the Zero-3 partitioning scheme can exhibit "super-linear" scaling with respect to the node count, as more aggregate bandwidth becomes available at increasing node count~\cite{rajbhandari2020zero}.
Generally, the runtime per epoch exhibited strong scaling, suggesting that hardware utilization remained steady with increasing node count, i.e., we did not reach a point of over-parallelization of the hardware at 16 nodes.


Given that the number of epochs increased with increasing node count under our experimental conditions, and the average power consumption increased in direct proportion to the node count, we found that the total energy consumed over the entire course of training increased with increasing node count. This underscores the need for individual users to consider the trade-offs between energy consumption during training and the overall time to solution, in cases where the runtime is strongly influenced by the algorithmic details of the problem. 
In other words, executing the same fine-tuning workload while increasing the number of nodes without selecting appropriate hyperparameters might not result in the expected time- and energy-efficiency gains.
Scaling these insights across the whole data center can drastically affect energy use, facility utilization and the effectiveness of scheduling strategies.

\subsection{Stable Diffusion (Image Generation) Training Workload Power Profiles}\label{sec:exp_finetuning_SF}

\begin{figure}[ht]
\centering
    \includegraphics[width=\linewidth]{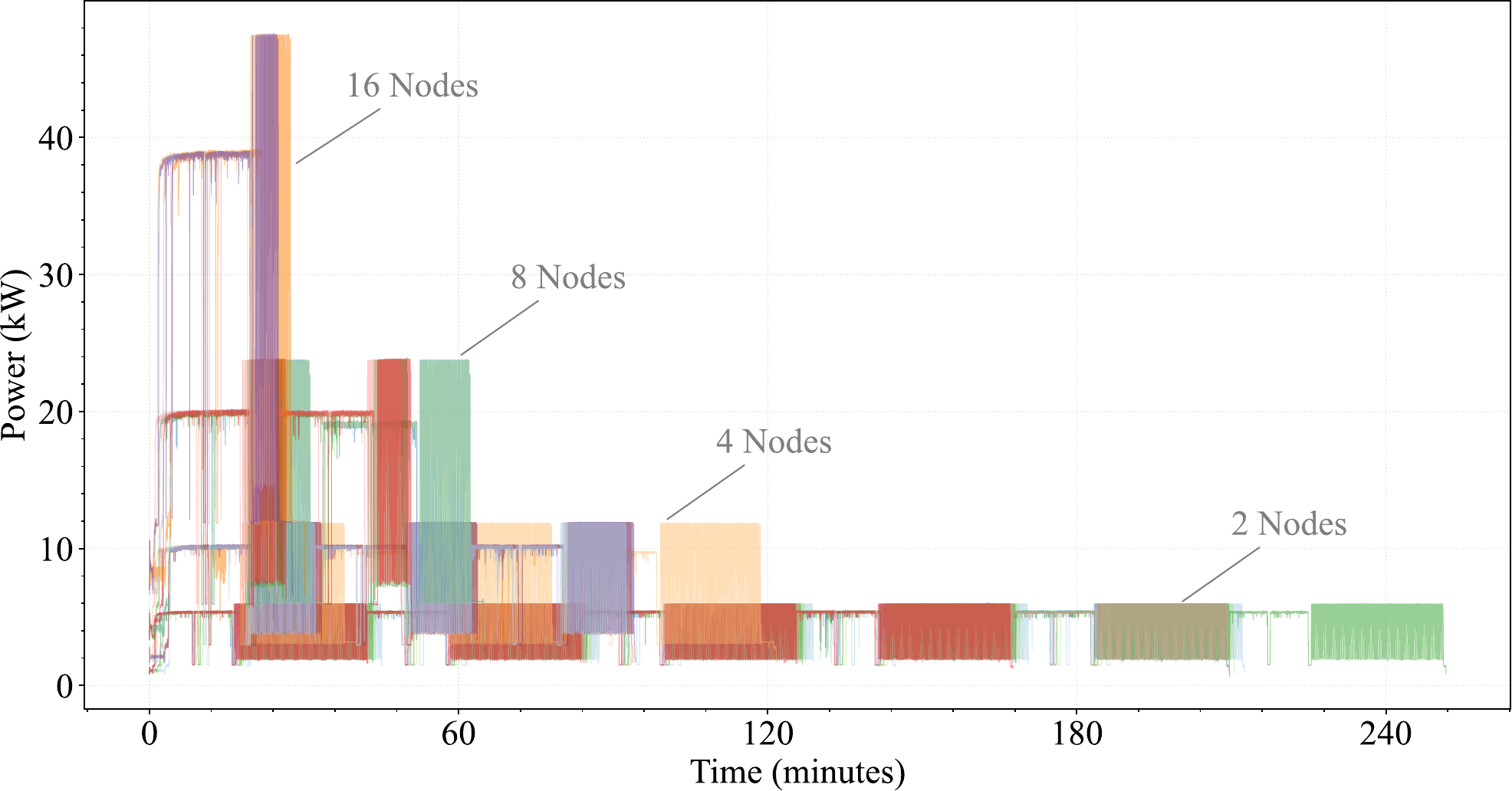}
    \caption{Stable Diffusion training power profiles. Replicate experiments using the same node count are shown as superimposed transparent curves with different colors. Each power profile aggregates power consumption across all CPU and GPU devices utilized by the workload.}
    \label{fig:stable_diffusion_training_power_profile}
\end{figure}

A similar set of results was generated for the Stable Diffusion image-generation model. The time-resolved power consumption profiles with the number of nodes are presented in Figure~\ref{fig:stable_diffusion_training_power_profile}.
A few key differences can be promptly identified, compared to the LLM fine-tuning results.
First, execution times were longer, particularly at lower node counts, taking up to 4.2 hours to complete.
Evaluation is also much more time-intensive for diffusion models due to the nature of diffusion sampling, which is iterative and expensive and requires many denoising steps for each sample, compared to LLM evaluation which can be a simple forward pass. 
Second, we observed a larger range of power draw over the course of training for a given node count compared to the Llama-2 case, particularly during the evaluation phases of the benchmark. In the 16-node configuration, power draw fluctuated between approximately 12~kW and 48~kW, reflecting alternating periods of reduced activity and short-duration, high-intensity compute associated with image generation.
These fluctuations were consistent with the diffusion sampling process used during evaluation, which involves repeated denoising steps and intermittent synchronization across devices.


\begin{figure}[ht]
    \begin{subfigure}{0.49\textwidth}
        \centering
        \includegraphics[width=\linewidth]{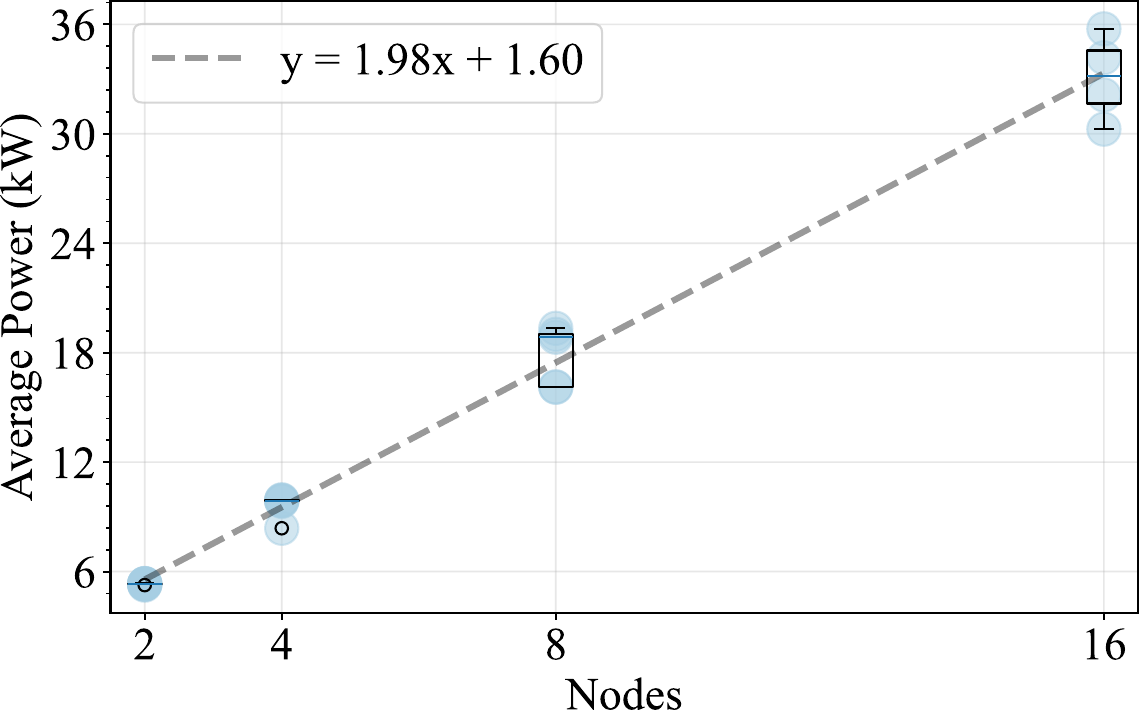}
        \caption{}
        \label{fig:stable_diffusion_subplots_power_avg}
    \end{subfigure}
    \begin{subfigure}{0.50\textwidth}
        \centering
        \includegraphics[width=\linewidth]{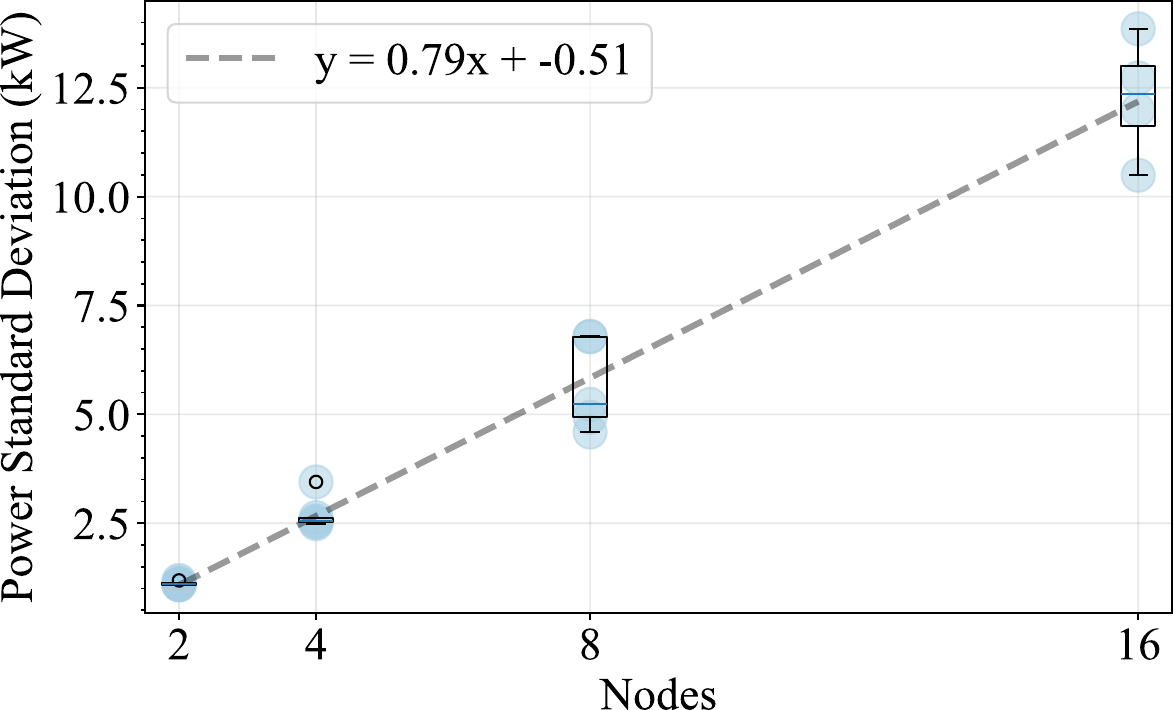}
        \caption{}
        \label{fig:stable_diffusion_subplots_power_std}
    \end{subfigure}
    \begin{subfigure}{0.495\textwidth}
        \centering
        \includegraphics[width=\linewidth]{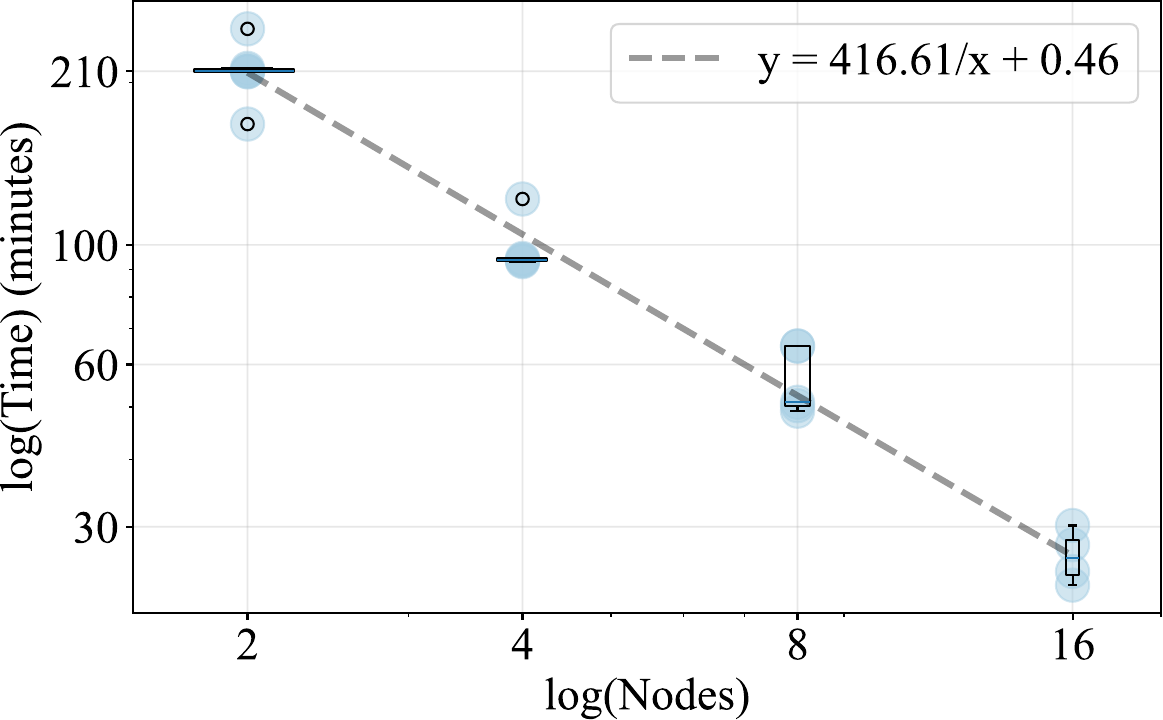}
        \caption{}
        \label{fig:stable_diffusion_subplots_time}
    \end{subfigure}
    \begin{subfigure}{0.49\textwidth}
        \centering
        \includegraphics[width=\linewidth]{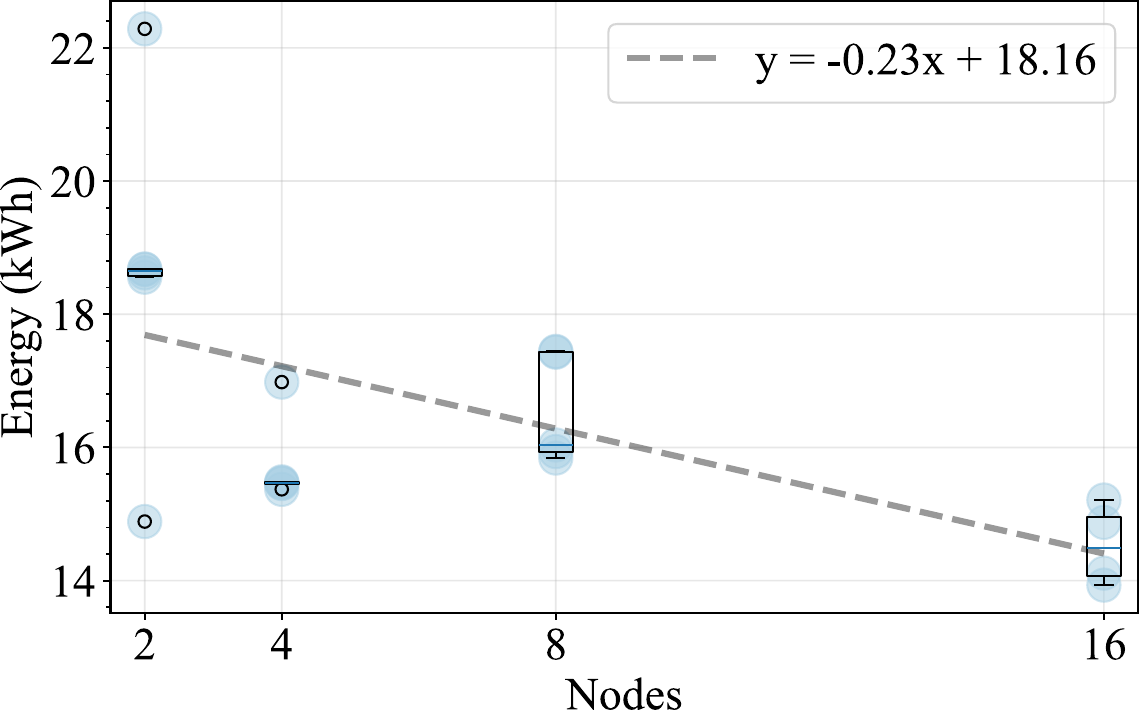}
        \caption{}
        \label{fig:stable_diffusion_subplots_energy}
    \end{subfigure}
    \caption{Stable Diffusion training energy metrics as a function of number of nodes. Replicate experiments are shown as red dots, vertically distributed with a box-and-whisker plot for each node value. Fitted curves are also presented as dashed lines.}
    \label{fig:stable_diffusion_subplots}
\end{figure}

Power-related metrics with node count for the Stable Diffusion image-generation training workload are presented in Figure~\ref{fig:stable_diffusion_subplots}.
Similar to the Llama-2 case, average power (Subplot~\ref{fig:stable_diffusion_subplots_power_avg}) and power variability (Subplot~\ref{fig:stable_diffusion_subplots_power_std}) increased linearly with node count, despite the clear visual differences in the time-resolved power profiles shown in Figures~\ref{fig:llama2_70b_lora_training_power_profile} and~\ref{fig:stable_diffusion_training_power_profile}. This highlights the need for time-resolved power profiles to inform data center power consumption analysis, as aggregate metrics may obscure the rapid but large fluctuations in the power drawn by the facility's hardware as these models run.

\begin{table}[ht]
    \centering
    \caption{Summary of runtimes from six replicate Stable Diffusion training workloads.}
        \begin{tabular}{lrrrrr}
        \label{tab:SDTrainingTimePerEpoch}
        \textbf{Nodes} & \textbf{\thead{Mean Total\\Runtime (sec)}}  & \textbf{\thead{Mean Epoch\\Runtime (sec)}} & \textbf{\thead{Mean Epoch\\Speedup}} & \textbf{\thead{Number of\\Epochs}} & \textbf{\thead{Number of\\Evaluations}} \\
        \hline
        1 & 24,632 ± 120 & 62,669 ± 306 & 0.00 & 0.39 ± 0.00 & 5.00 ± 0.00 \\
        2 & 12,480 ± 1,576 & 31,751 ± 52 & 1.97 & 0.39 ± 0.05 & 5.00 ± 0.63 \\
        4 & 5,784 ± 792 & 12,262 ± 1,679 & 2.59 & 0.47 ± 0.00 & 3.00 ± 0.00 \\
        8 & 3,202 ± 421 & 5,091 ± 669 & 2.41 & 0.63 ± 0.00 & 2.00 ± 0.00 \\
        16 & 1,432 ± 149 & 2,277 ± 238 & 2.24 & 0.63 ± 0.00 & 1.00 ± 0.00 \\
        \hline
        \end{tabular}
\end{table}

Our analysis of the factors impacting runtime in the Llama-2 case applied similarly to the Stable Diffusion case. In the Stable Diffusion case, evaluation occurred every 50 training samples (rather than 48 for Llama-2), and the ratio between evaluation time and weight updating time shown in Table~\ref{tab:SDTrainingTimePerEpoch} was much larger than in the Llama-2 case. Since we kept the number of steps between evaluations the same for all training runs, and the global batch size increased proportionally with the number of nodes, we saw the smaller nodes spend significantly more time in each evaluation phase and also have more total evaluation phases. 

As a result, total energy consumption, shown in Subplot~\ref{fig:stable_diffusion_subplots_energy}, tended to decrease as node count increases, indicating that the reduction in execution time shown in Subplot~\ref{fig:stable_diffusion_subplots_time} offset the total energy consumed from the addition of more nodes. We note that this is not a generalizable result, as it was the result of the aforementioned details concerning the evaluations. The observed variability in power draw changed significantly between the two models, highlighting the potential impact of different model architectures on facility-level load profiles.


\subsection{Llama-3 70B (LLM) Offline Inference Workload Power Profiles}

\begin{figure}[ht]
\centering
    \includegraphics[width=\linewidth]{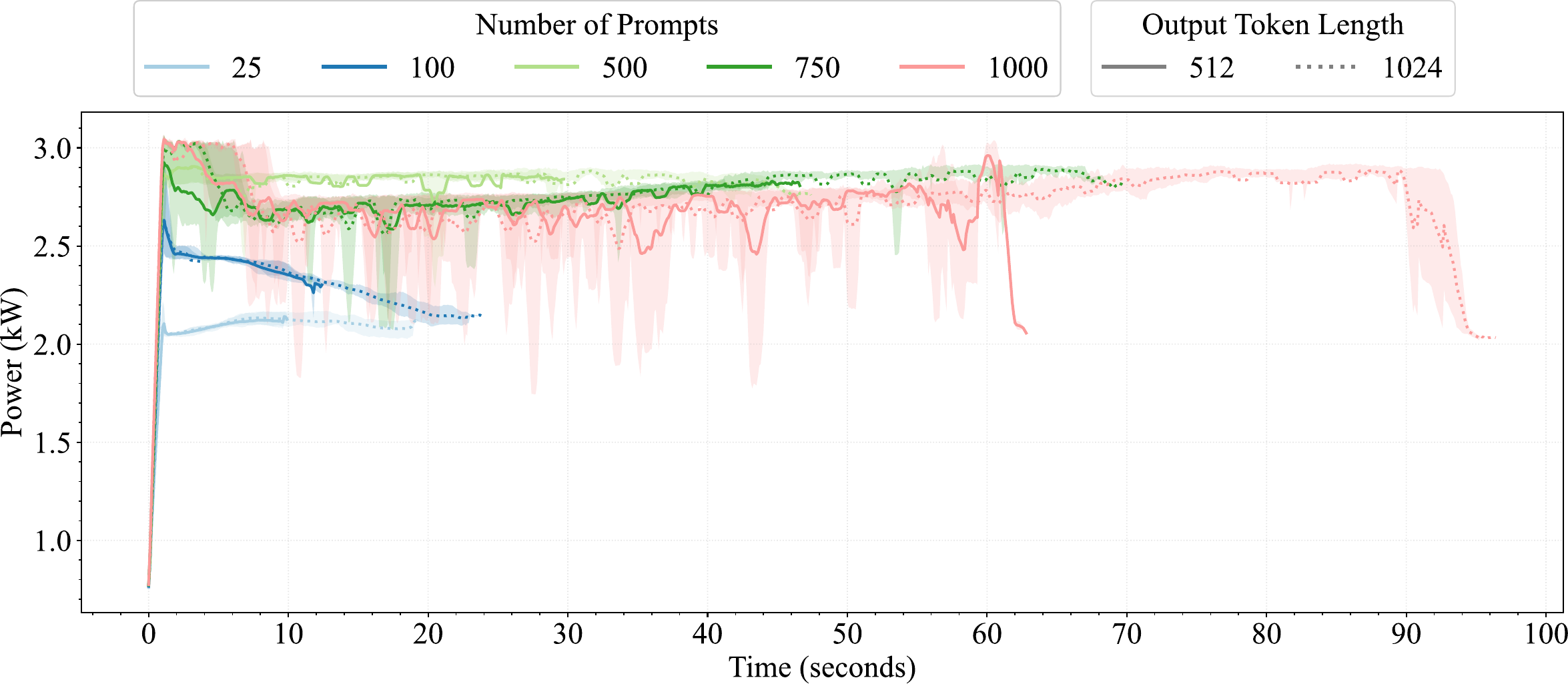}
    \caption{Llama-3 70B offline inference power profiles as a function of number of prompts and output token length. 
    Median values are shown as lines, and 10-90$^{th}$ percentiles as shaded regions.}
    \label{fig:inference_offline_power_profile}
\end{figure}

In this section, offline inference benchmarks for the Llama-3 70B model were evaluated using the vLLM framework on one compute node.
Figure~\ref{fig:inference_offline_power_profile} shows the effect of varying the max token output values and the number of input prompts on the time-resolved power consumption profiles.
A very fast ramp-up period was observed across all scenarios, followed by either a sustained or decreasing trend.
The overall execution time was in the order of seconds, ranging from $\sim$10 seconds for the shortest case to $\sim$85 seconds for the longest.
The max number of output tokens seemed to proportionately impact execution time, i.e., doubling the number of max output tokens roughly doubled the execution time.
In contrast, power consumption increased with the number of input prompts, but only up to a certain point (around 325 input prompts), saturating at 2.8 kW after that.
Curves with high number of input prompts also exhibited a higher variance across replicates, with a few curves presenting pronounced intermittent dips.

\begin{figure}[ht]
    \centering
    \begin{subfigure}{0.25\textwidth}
        \centering
        \includegraphics[width=\linewidth]{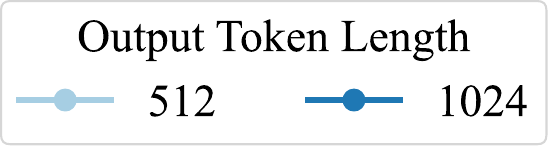}
    \end{subfigure}
    \\
    \vspace{5pt}
    \begin{subfigure}{0.47\textwidth}
        \centering
        \includegraphics[width=\linewidth]{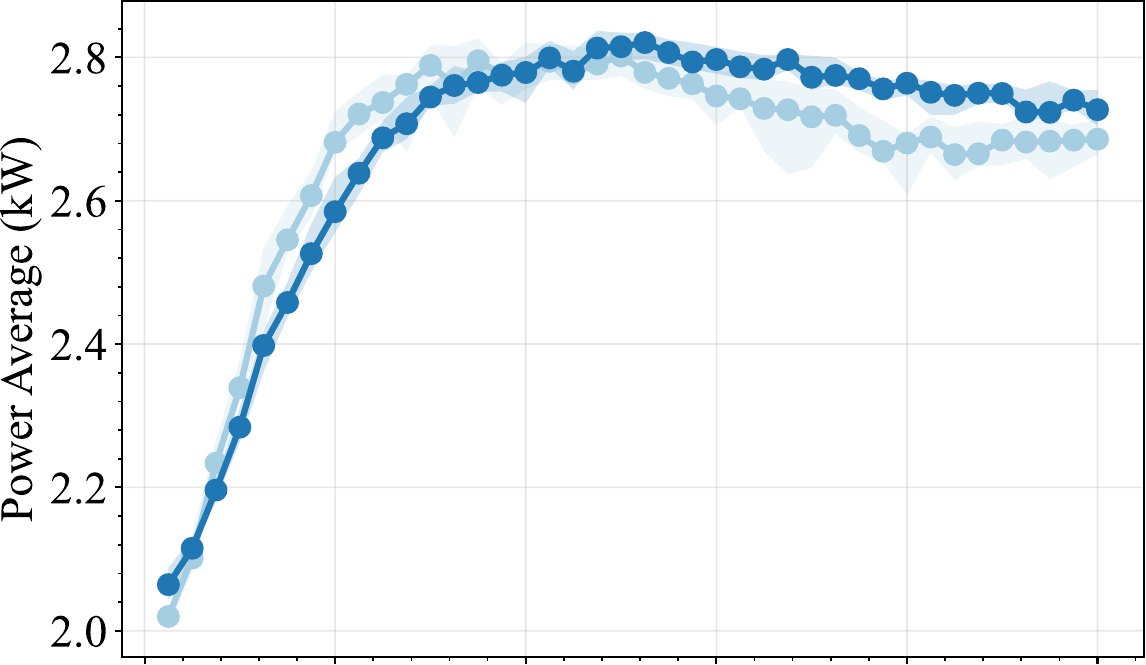}
        \caption{}
        \label{fig:llama2_70b_lora_inference_subplots_power_avg}
    \end{subfigure}
    \begin{subfigure}{0.48\textwidth}
        \centering
        \includegraphics[width=\linewidth]{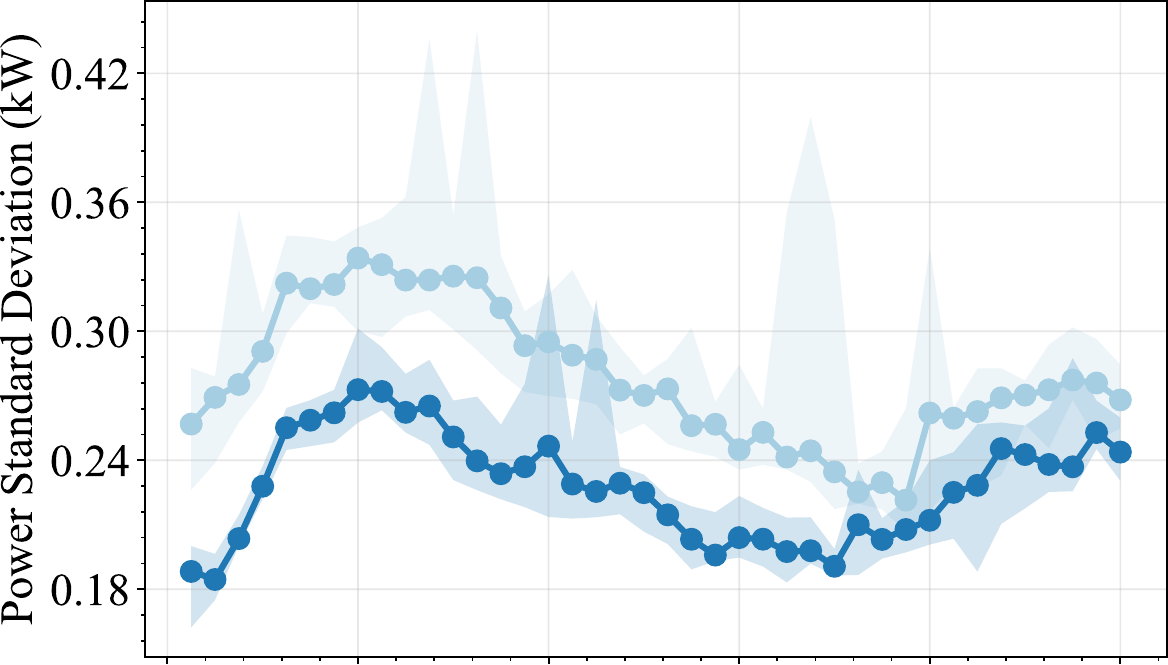}
        \caption{}
        \label{fig:llama2_70b_lora_inference_subplots_power_std}
    \end{subfigure}
    \begin{subfigure}{0.47\textwidth}
        \centering
        \includegraphics[width=\linewidth]{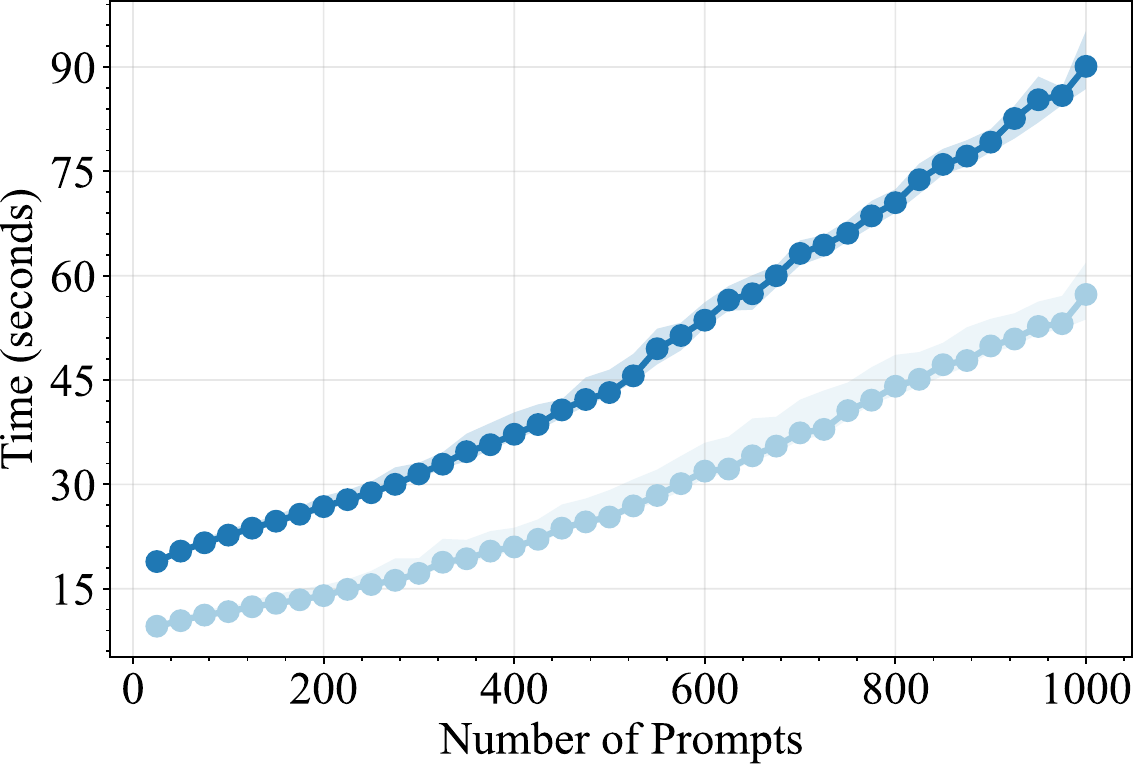}
        \caption{}
        \label{fig:llama2_70b_lora_inference_subplots_time}
    \end{subfigure}
    \begin{subfigure}{0.48\textwidth}
        \centering
        \includegraphics[width=\linewidth]{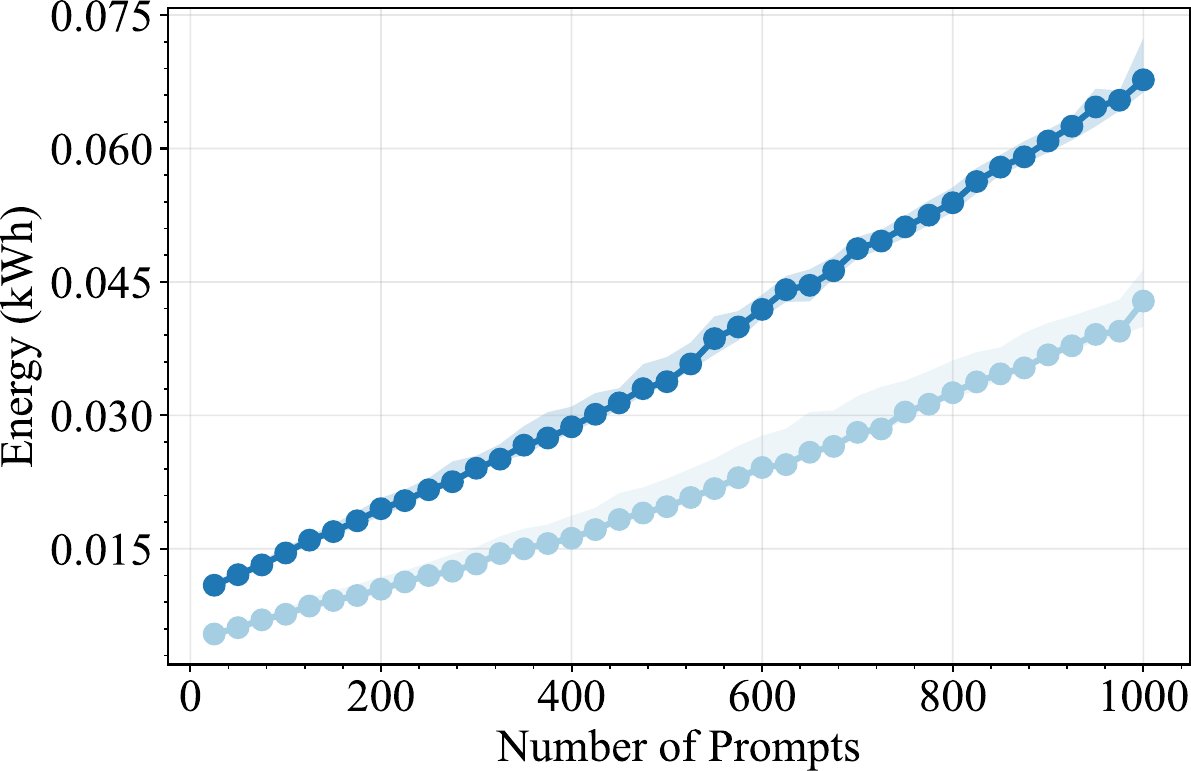}
        \caption{}
        \label{fig:llama2_70b_lora_inference_subplots_energy}
    \end{subfigure}
    \caption{Llama-3 70B offline inference energy metrics as a function of number of prompts and output tokens. Median values across experiments are shown as dots, and 10-90$^{th}$ percentiles as a shaded regions.}
    \label{fig:llama2_70b_lora_inference_subplots}
\end{figure}

Power-related metrics are presented in Figure~\ref{fig:llama2_70b_lora_inference_subplots} as a function of number of prompts and of max output tokens.
Subplot~\ref{fig:llama2_70b_lora_inference_subplots_power_avg} presents the average power consumption, highlighing the saturating trend with the number of prompts.
Surprisingly, average power peaked at an intermediary value of input prompts between 300 and 500, above which it decreased slightly. 
Power fluctuations remained mostly constant, as shown in Subplot~\ref{fig:llama2_70b_lora_inference_subplots_power_std}, a modest peak also being achieved at a low range of input prompt size.
Execution time, shown in Subplot~\ref{fig:llama2_70b_lora_inference_subplots_time}, increased with number of input prompts in a near piecewise linear fashion, the slope changing at around 450 prompts.
The total energy consumption, shown in Subplot~\ref{fig:llama2_70b_lora_inference_subplots_energy}, was once more calculated from the average power and execution time, and exhibited a more linear behavior. 
The number of max output tokens seemed to affect power variability, execution time, and energy proportionally, but average power consumption in a more subtle and less intuitive way.


\subsection{Online/Serve Inference Workload Power Profiles}\label{sec:exp_online_inference}

\begin{figure}[ht]
    \centering
    \begin{subfigure}{0.8\textwidth}
        \centering
        \includegraphics[width=\linewidth]{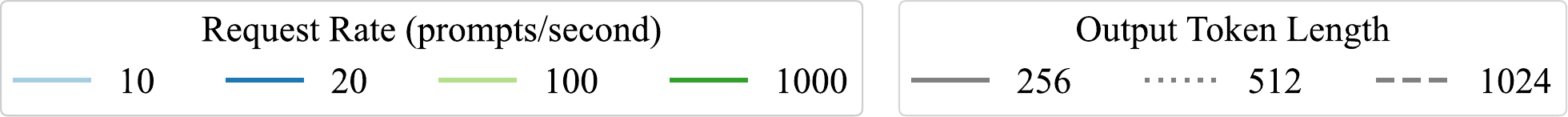}
    \end{subfigure}
    \\
    \vspace{5pt}
    \begin{subfigure}{0.52\textwidth}
        \centering
        \includegraphics[width=\linewidth]{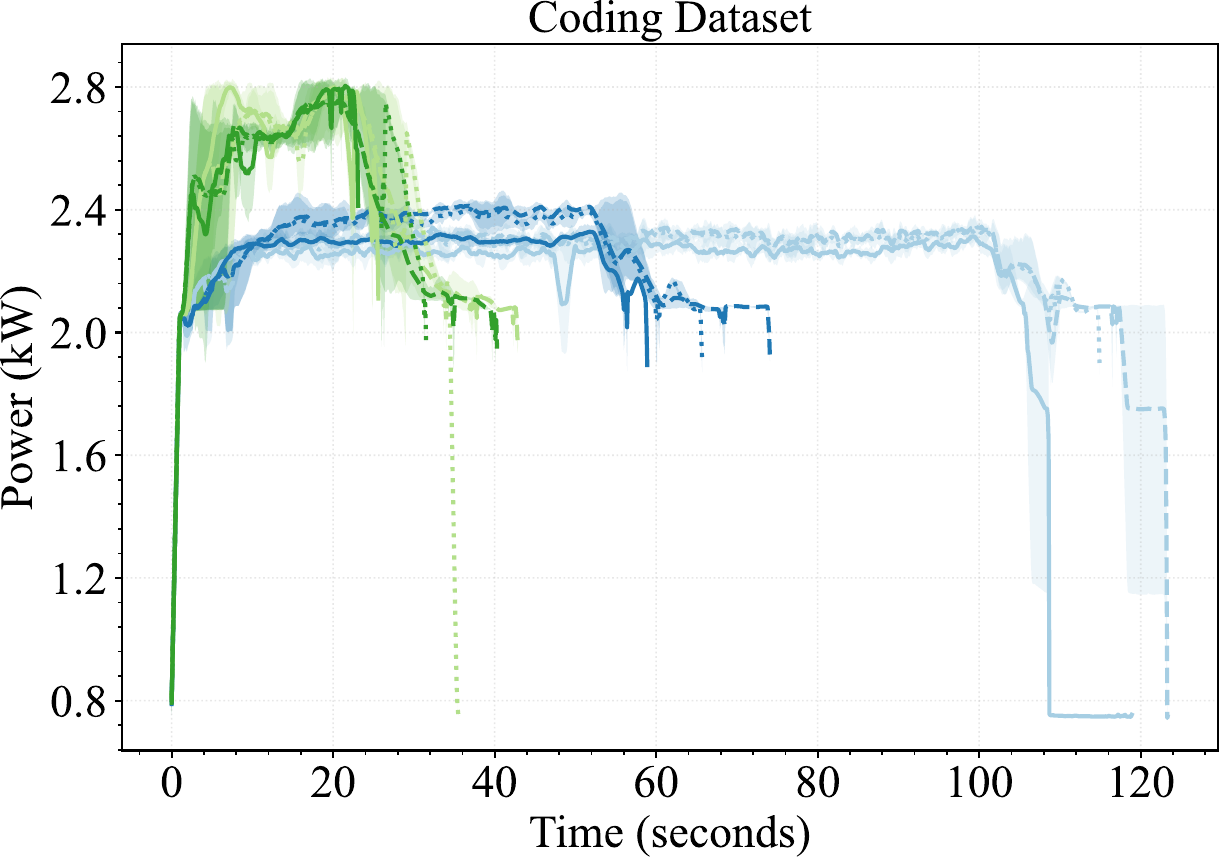}
        \caption{}
    \end{subfigure}
    \begin{subfigure}{0.47\textwidth}
        \centering
        \includegraphics[width=\linewidth]{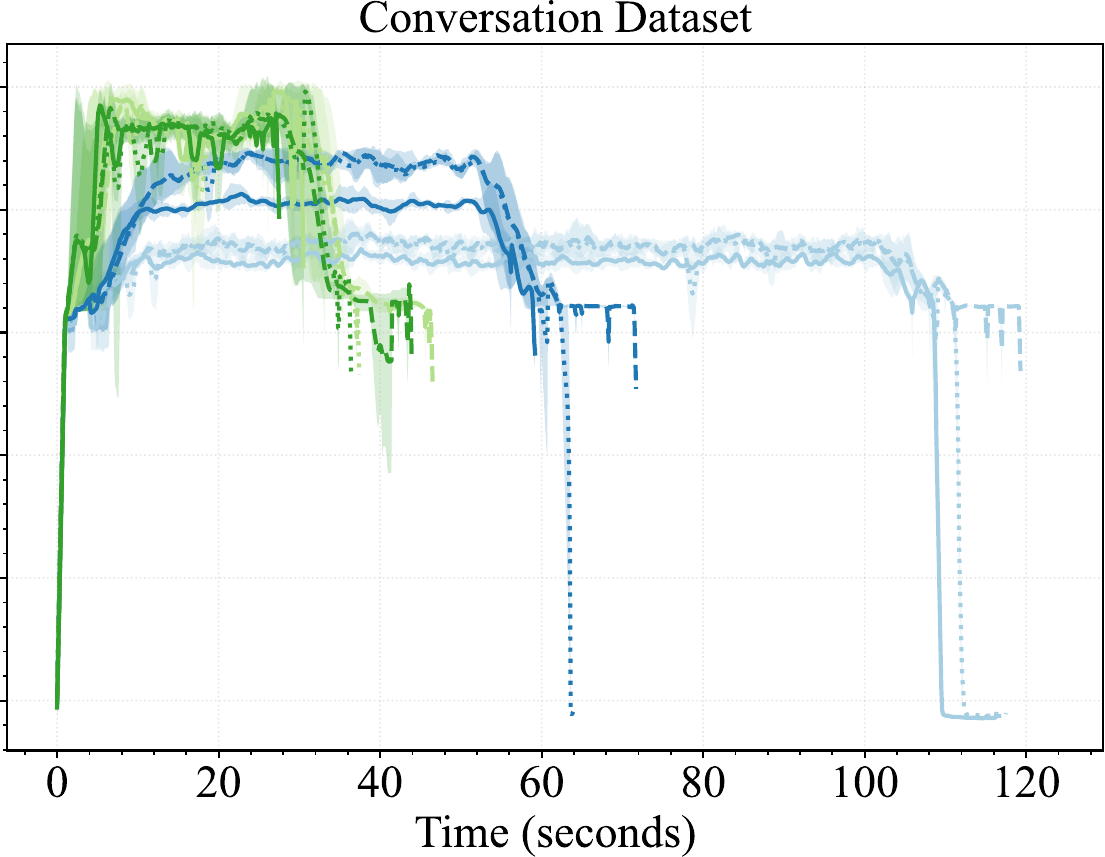}
        \caption{}
    \end{subfigure}
    \caption{Power series of Llama-3 70B online inference benchmark as a function of dataset, request rate and output token length, assuming a constant batch size of 1,000 prompts.
    Median values are shown as lines, and 10-90$^{th}$ percentiles as shaded regions.}
    \label{fig:online_inference_series}
\end{figure}

In the previous section, we benchmarked LLM inference in offline/batch mode, i.e. submitting all prompts at once.
Instead, here we present results from testing  the Llama-3 70B model in online/serve mode -- serving the model on one node and processing streams of requests -- which more closely resembles a commercial inference operation. See Section~\ref{sec:met-inference} for more details on the difference between offline/batch and online/serve operation modes.
We tested prompts from two datasets: \textit{likaixin/InstructCoder}, for code completion and editing, and \textit{mgoin/mlperf-inference-Llama-2-data}, with conversation prompts \cite{li2024instructcoder, mgoing2025}. 
We set the total number of requests to 1,000 prompts, and ran tests varying the dataset, the request rate, the output token length and the seed. 
For each experiment we executed three repetitions using different seeds.

Figure~\ref{fig:online_inference_series} shows the power consumption timeseries for a select number of request rates. 
The profiles were fairly consistent across datasets, and were mostly influenced by the request rate.
Some effect due to output token length can be observed in the tail of the profiles.
Interestingly, there were more pronounced differences between tests assuming 10 and 20 prompts per second, than between 100 and 1,000, indicating an early saturation in the system's ability to process requests.

\begin{figure}[ht]
    \centering
    \begin{subfigure}{0.4\textwidth}
        \centering
        \includegraphics[width=\linewidth]{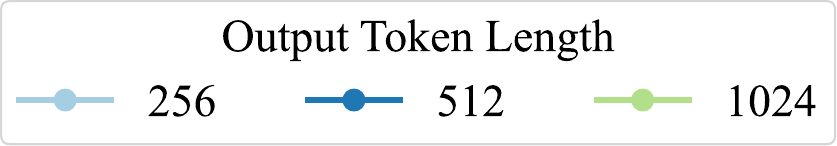}
    \end{subfigure}
    \\
    \vspace{5pt}
    \begin{subfigure}{0.525\textwidth}
        \centering
        \includegraphics[width=\linewidth]{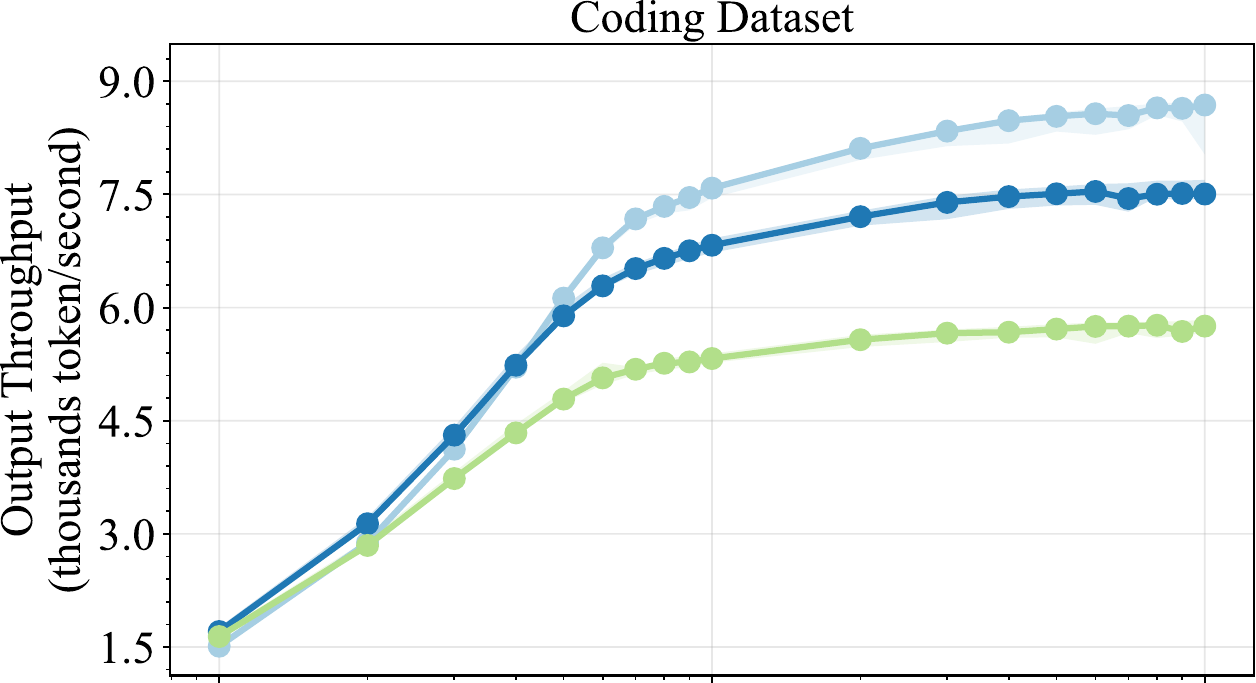}
        \caption{}
        \label{fig:online_inference_performance-a}
    \end{subfigure}
    \begin{subfigure}{0.46\textwidth}
        \centering
        \includegraphics[width=\linewidth]{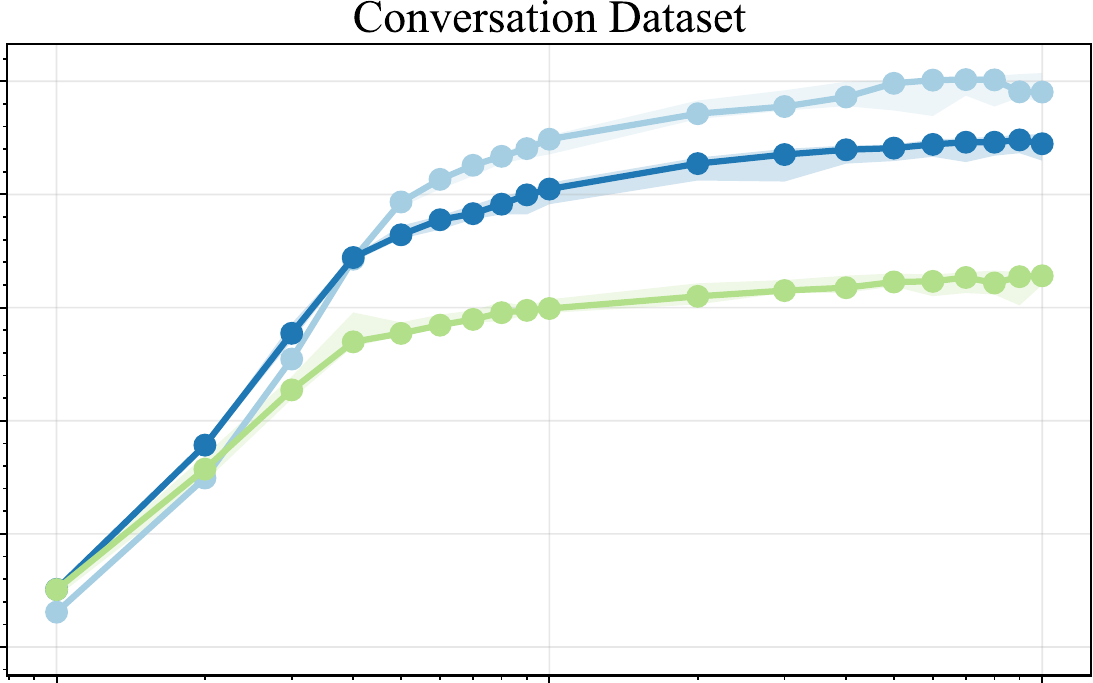}
        \caption{}
        \label{fig:online_inference_performance-b}
    \end{subfigure}
    \\
    \begin{subfigure}{0.525\textwidth}
        \centering
        \includegraphics[width=\linewidth]{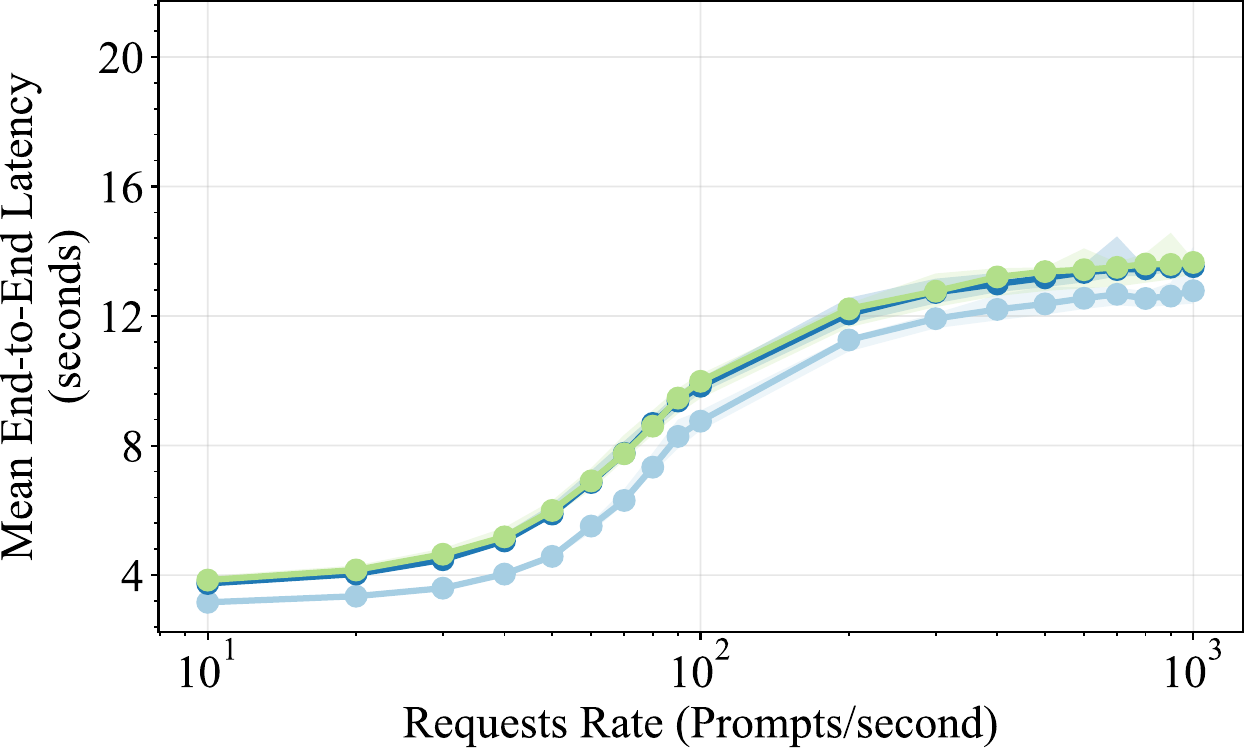}
        \caption{}
        \label{fig:online_inference_performance-c}
    \end{subfigure}
    \begin{subfigure}{0.46\textwidth}
        \centering
        \includegraphics[width=\linewidth]{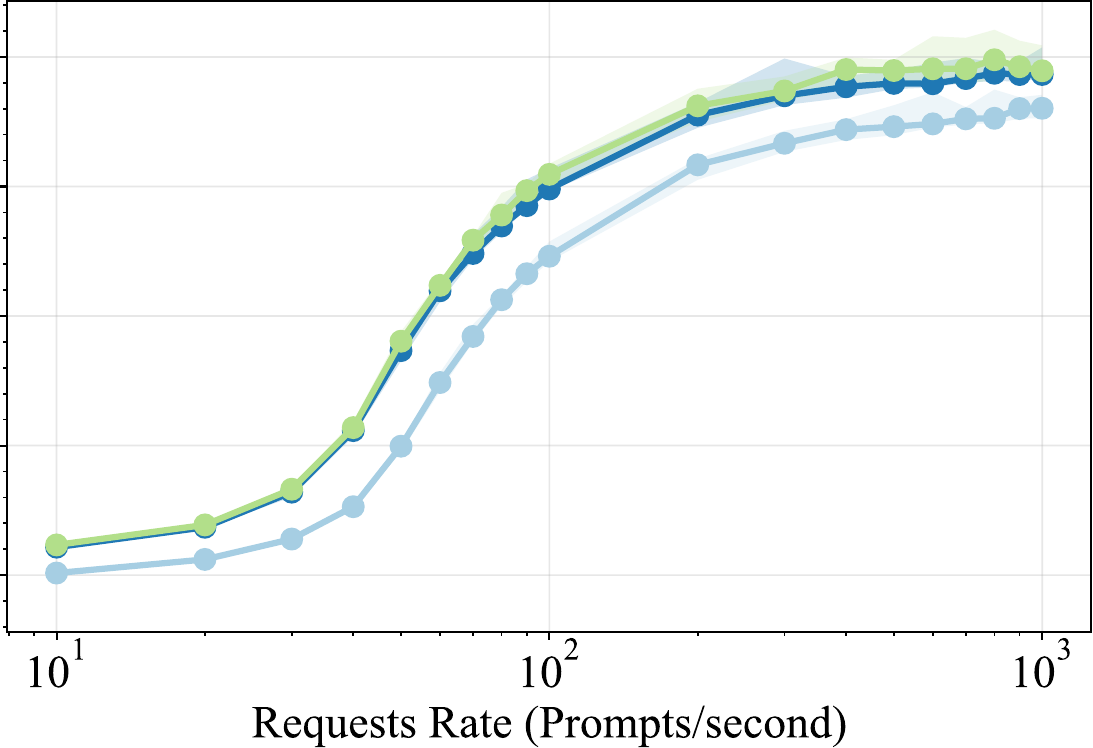}
        \caption{}
        \label{fig:online_inference_performance-d}
    \end{subfigure}
    \caption{Llama-3 70B offline inference performance metrics
    as a function of dataset, request-rate (in log-scale) and output token length.
    Median values are shown as lines, and 10-90$^{th}$ percentiles as shaded regions.
    }
    \label{fig:online_inference_performance}
\end{figure}

\begin{figure}[t]
    \centering
    \begin{subfigure}{0.38\textwidth}
        \centering
        \includegraphics[width=\linewidth]{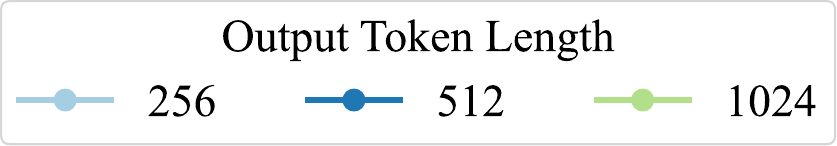}
    \end{subfigure}
    \\
    \vspace{5pt}
    \begin{subfigure}{0.47\textwidth}
        \centering
        \includegraphics[width=\linewidth]{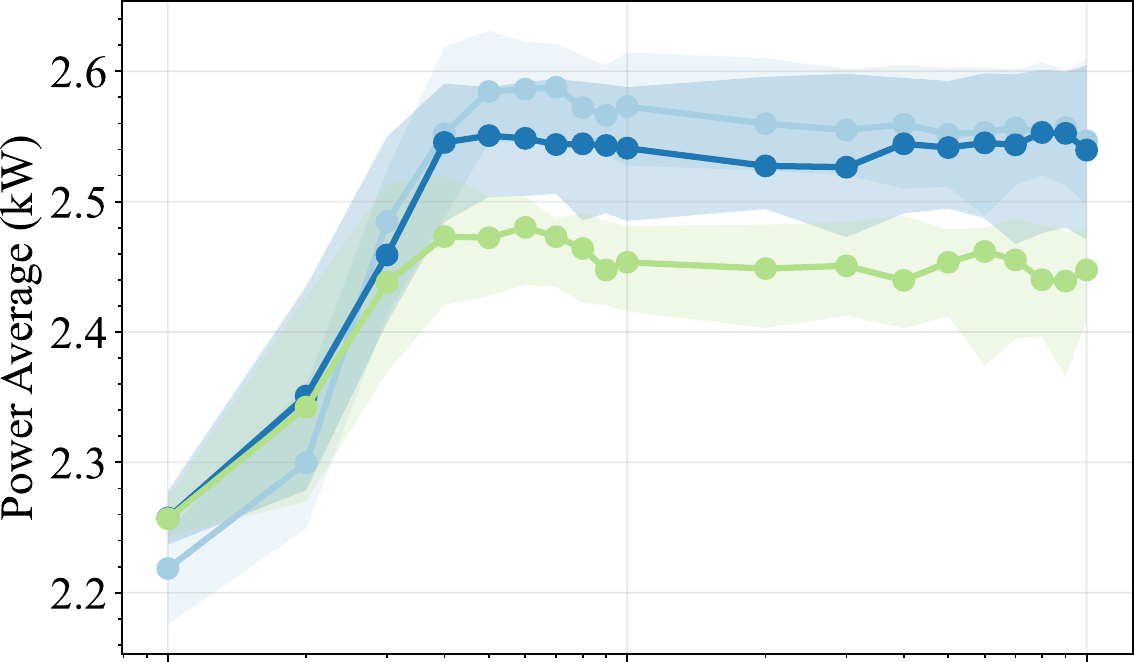}
        \caption{}
        \label{fig:online_inference_power_power_avg}
    \end{subfigure}
    \begin{subfigure}{0.48\textwidth}
        \centering
        \includegraphics[width=\linewidth]{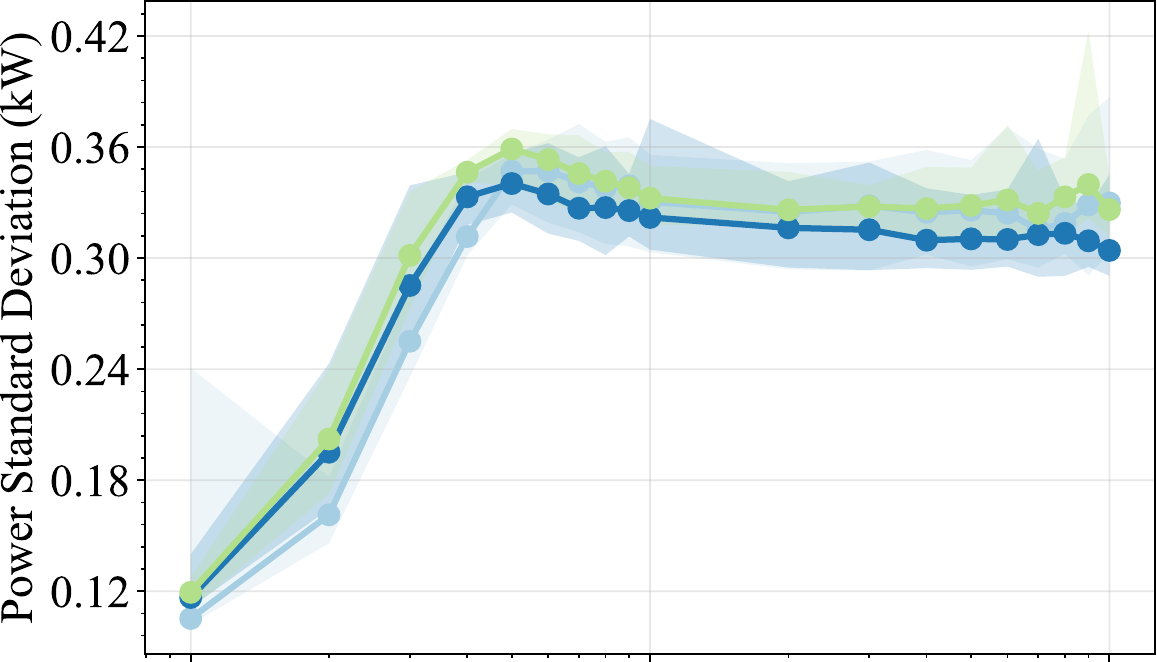}
        \caption{}
        \label{fig:online_inference_power_power_std}
    \end{subfigure}
    \begin{subfigure}{0.47\textwidth}
        \centering
        \includegraphics[width=\linewidth]{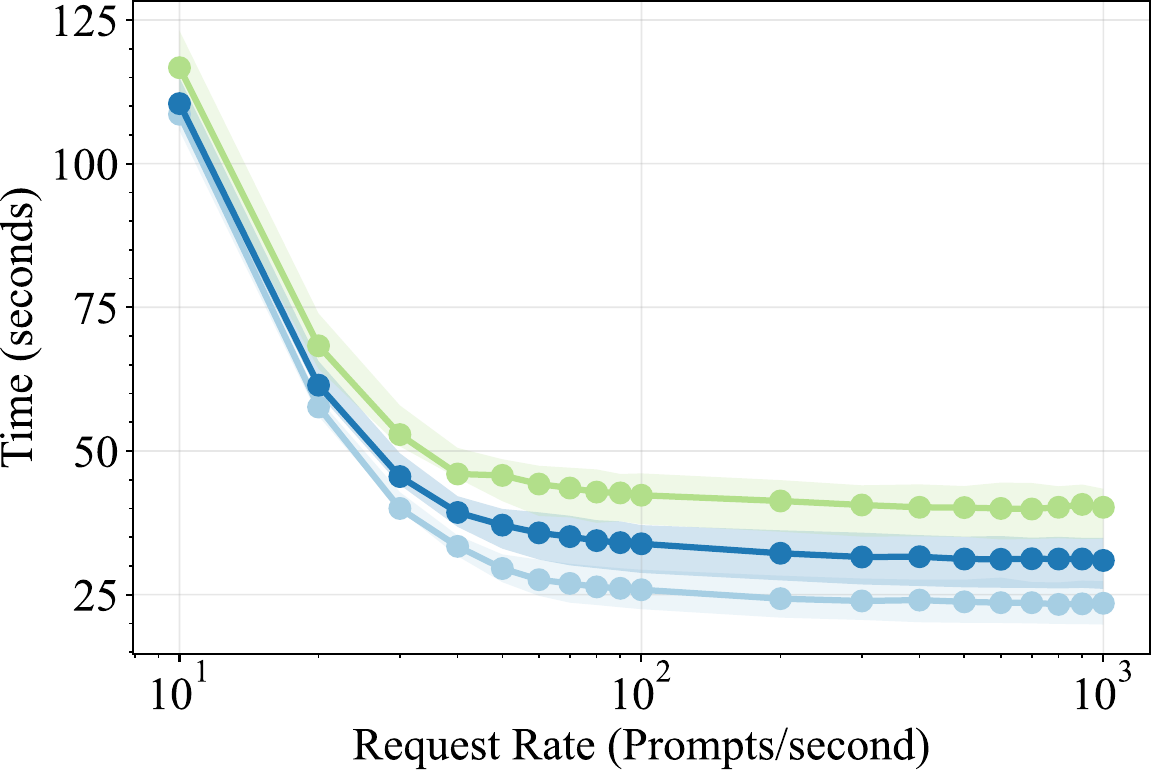}
        \caption{}
        \label{fig:online_inference_power_time}
    \end{subfigure}
    \begin{subfigure}{0.48\textwidth}
        \centering
        \includegraphics[width=\linewidth]{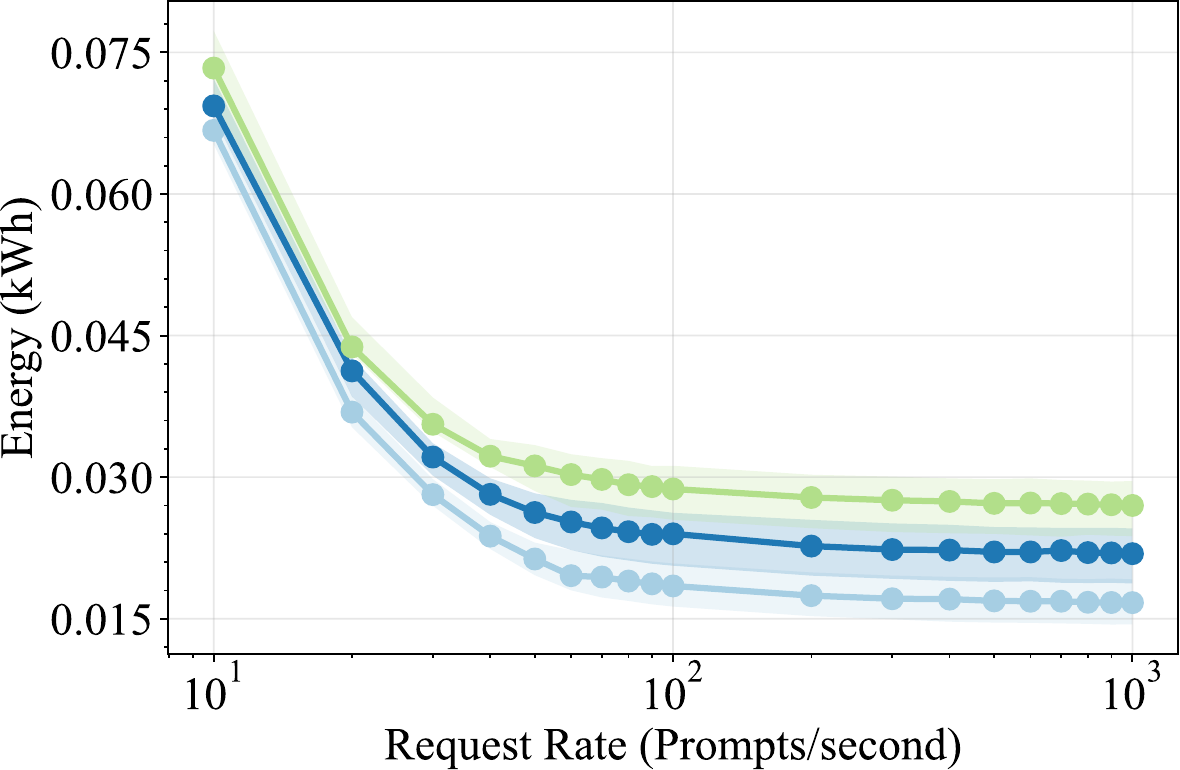}
        \caption{}
        \label{fig:online_inference_power_energy}
    \end{subfigure}
    \caption{Llama-3 70B online inference energy metrics as a function of dataset, request-rate (in log-scale) and output token length.
    Median values are shown as lines, and 10-90$^{th}$ percentiles as shaded regions.
    }
    \label{fig:online_inference_power}
\end{figure}

Figure~\ref{fig:online_inference_performance} shows tradeoffs in performance as a function of request rate (in log-scale).
The first row -- Figures~\ref{fig:online_inference_performance-a}-\ref{fig:online_inference_performance-b} -- shows average output throughput (in thousands of tokens per second), which as expected increased with request rate but leveled out as the  rate gets closer to 1,000 prompts per second, indicating the performance degradation associated with processing more prompts. 
We also noted how requested output token length had increasingly significant impact on throughput as the request rate increased.
Subplots~\ref{fig:online_inference_performance-c}-\ref{fig:online_inference_performance-d} in the second row present average end-to-end latency, which measures the average time between when the prompt is received and when the response is complete.
Performance degraded at a higher pace with lower request rates (between 1 and 100 prompts per seconds), while leveling off in the upper part of the test range (between 100 and 1,000 prompts per second).
Also, prompts from the conversation dataset in Subplot~\ref{fig:online_inference_performance-d} resulted in noticeably higher latency than coding prompts in Subplot~\ref{fig:online_inference_performance-c}.
Output token length seemed to have a smaller impact on this metric than on output throughput.
Latency is one of the most important metrics to quantify online inference performance, as it can be easily connected to user experience.
As discussed in Section~\ref{sec:model_structure}, in the whole-facility simulation we used latency as the metric correlating total incoming request rate at the inference data center to the number of model instances required.
This assumption affects facility performance, server utilization and power consumption.
None of the metrics were particularly impacted by seed selection.

Figure~\ref{fig:online_inference_power} shows power-related metrics as a function of dataset, request rate, and output token length.
Subplot~\ref{fig:online_inference_power_power_avg} presents the average power consumption during the tests, which rapidly increased to $\sim$2.9 kW with request rates less than 100 prompts/second; Subplot~\ref{fig:online_inference_power_power_std} shows the standard deviation on power consumption, which also similarly saturated rapidly.
Subplot~\ref{fig:online_inference_power_time} and~\ref{fig:online_inference_power_energy} present respectively the time to execute the test and the energy consumed. 
Both metrics rapidly decreased with higher request rates, but converged asymptotically, indicating that saturation in the performance of the setup impeded gains in processing prompts at higher request rates.
Comparing the latency subplots in Figure~\ref{fig:online_inference_performance-c} and~\ref{fig:online_inference_performance-d} and  Figure~\ref{fig:online_inference_power_energy}), we note an interesting tradeoff that should inform resources allocation at the data center level: processing a fixed number of prompts on fewer model instances results in a higher effective request rate on each model instance, which is more energy-efficient but will result in a degraded user experience (i.e. a higher latency).

\begin{figure}[ht]
    \centering
    \begin{subfigure}{0.5\textwidth}
        \centering
        \includegraphics[width=\linewidth]{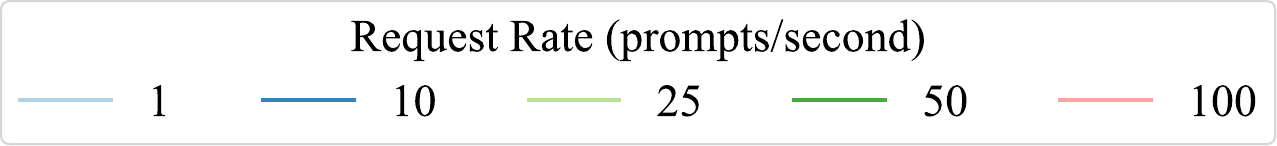}
    \end{subfigure}
    \\
    \vspace{5pt}
    \begin{subfigure}{0.52\textwidth}
        \centering
        \includegraphics[width=\linewidth]{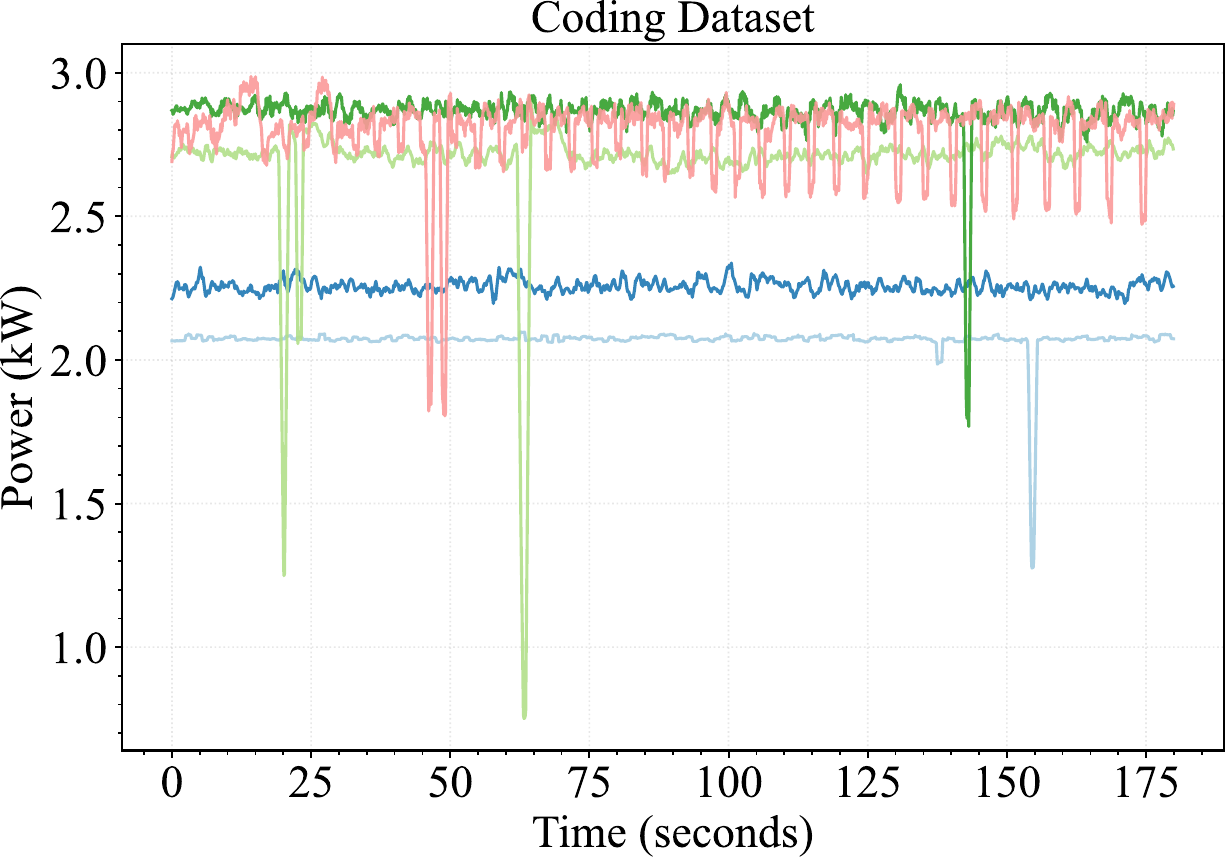}
        \caption{}
    \end{subfigure}
    \begin{subfigure}{0.47\textwidth}
        \centering
        \includegraphics[width=\linewidth]{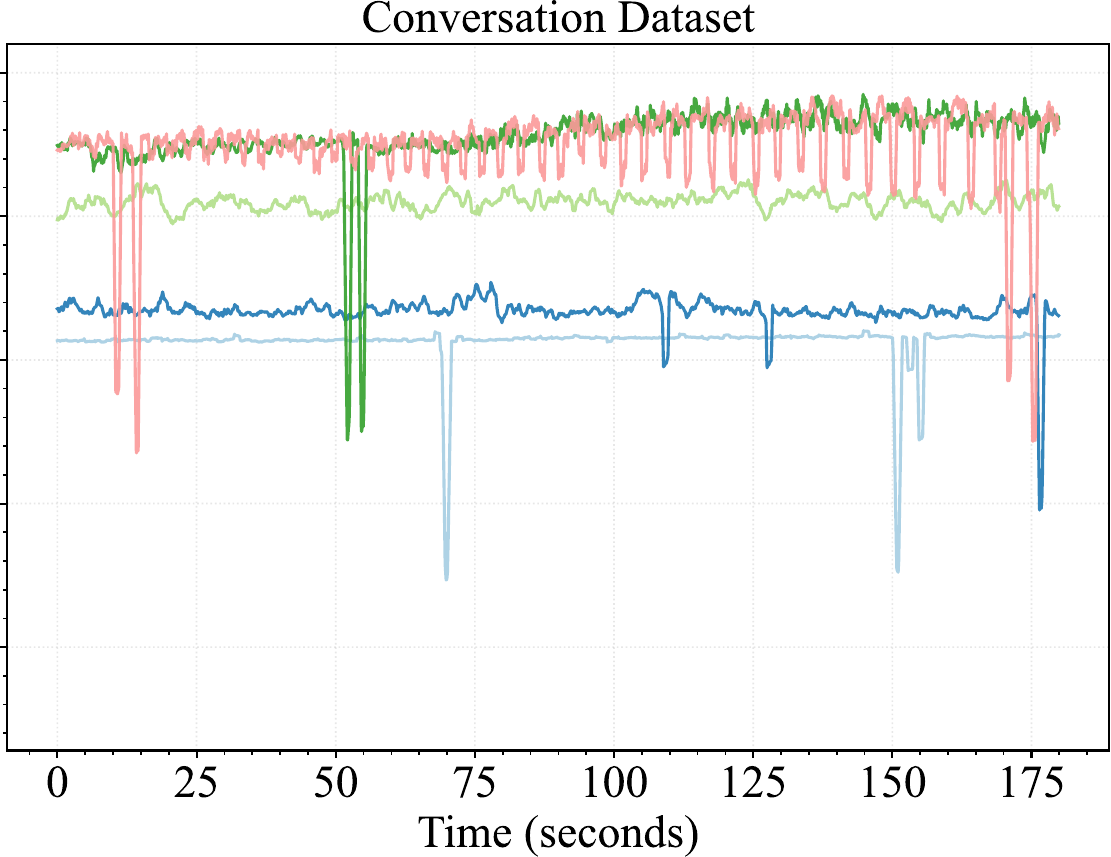}
        \caption{}
    \end{subfigure}
    \caption{Power series of Llama-3 70B online inference benchmark as a function of dataset, request rate and output token length, assuming a sustained request rate for 3 minutes.}
    \label{fig:online_inference_series_rate}
\end{figure}

While the online inference benchmarks in Figures~\ref{fig:online_inference_series},~\ref{fig:online_inference_performance} and~\ref{fig:online_inference_power} used a finite batch of prompts to measure performance, the whole-facility simulation model DIPLOEE assumes the data center processes a continuous stream of prompts characterized by a request rate that varies in time as a function of user-behavior.
Therefore, we executed one additional test, sampling power consumption during model inference while sustaining selected request rates over 3 minutes.
Snippets of these power samples are presented in Figure~\ref{fig:online_inference_series_rate}, and showed similar patterns between coding and conversation datasets, with average consumption saturating near $\sim$2.8~kW at 25 prompts/second, and sudden drops in consumption during the test. 
When simulating model instances processing a certain request rates, we used these power samples in DIPLOEE to define the server power consumption -- see Section~\ref{sec:DIPLOEE_inference} for additional details.

\section{Whole-Facility Simulation Results}
\label{sec:case_study}

We present two case studies using the methodology described in Section~\ref{sec:model_structure}.
In Section~\ref{sec:wf_results_colocation}, we simulated the operation of a 10 MW colocation data center running a mix of fine-tuning and training jobs.
In Section~\ref{sec:wf_results_inference}, we simulated operation of a 1 MW inference data center serving inference models in production.

The literature suggests considerable variability in data center size, measured in both the number of servers and overall power capacity, across different application. For the initial use case, facility sizing was informed by typical characteristics of colocation data centers. Such facilities generally exhibit fewer siting constraints, are significantly smaller than hyperscale data centers -- which may exceed 100 MW in capacity -- and typically operate within a range of 500 kW to 20 MW.

By contrast, AI inference facilities are commonly sited in close proximity to end users and within urban environments to minimize latency. These geographical constraints limit available power supply and thus the achievable facility size relative to colocation centers. While the selected sizes are intended to be representative, the DIPLOEE framework maintains flexibility to simulate the operation of data centers across a wide range of capacities.

For each of the use cases, we simulated operation for one full year using a one minute timestep.
We executed four separate simulations assuming four different average target utilization values, 20, 40, 60 and 80\% respectively.
Refer to Equation~\ref{eq:utilization} for how utilization is defined in this context.
We assumed the same node architecture as NLR's HPC: two AMD EPYC 9554 (Genoa) CPUs rated at 360 W, and 4 NVIDIA H100 SXM GPU accelerators rated at 700 W each, for a total of 3.520 kW per node.
The idle power for each node, which is the power consumed when the node is not actively used, was set to 420 W in accordance with the tests in Section~\ref{appA}.


\subsection{Colocation Data Center}\label{sec:wf_results_colocation}

We simulated operation at a 10 MW colocation data center running a mix of AI training and fine-tuning jobs.
The single job data, including power consumption, was taken from the experimental measurements obtained in Sections~\ref{sec:exp_finetuning_LL} and~\ref{sec:exp_finetuning_SF}, Llama-2 70B fine-tuning and Stable Diffusion training workloads, respectively.
The facility utilization distributions, including hourly, day of the week, job type and node count were adapted from the job data measured at Shanghai AI Lab data center used to develop new AI models~\cite{acme_paper, acme_dataset}.
Since this data did not span a whole year, we also used monthly distributions from NLR's Kestrel HPC to account for seasonality in facility utilization.
We simulated operation in a data center sized for 10 MW not including cooling and auxiliary loads. 
Assuming the same server architecture as NLR's Kestrel, such a facility  could support approximately 2,840 nodes.

\begin{table}[ht]
\caption{Colocation data center aggregate metrics across the whole year for different target utilization levels.}
\small
\centering
\begin{tabular}{lcccc}
& \multicolumn{4}{c}{\textbf{Average Utilization}} \\
\hline
 & 20\% & 40\% & 60\% & 80\% \\
\hline
\multicolumn{5}{l}{\textbf{Power}} \\
\quad Mean (MW) & 2.38 & 3.56 & 4.74 & 5.91 \\
\quad Std Dev (MW) & 0.85 & 1.59 & 1.91 & 1.78 \\
\quad Median (MW) & 2.22 & 3.27 & 4.69 & 7.00 \\
\quad 90th Percentile (MW) & 3.53 & 6.40 & 7.08 & 7.12 \\
\quad Max (MW) & 6.08 & 7.34 & 7.31 & 7.32 \\
\quad Peak-to-Avg Ratio & 2.56 & 2.06 & 1.54 & 1.24 \\
\hline
\multicolumn{5}{l}{\textbf{Utilization}} \\
\quad Mean (\%) & 20.14 & 40.24 & 60.34 & 80.34 \\
\quad Std Dev (\%) & 14.47 & 27.14 & 32.54 & 30.29 \\
\quad Median (\%) & 17.39 & 35.28 & 59.51 & 99.93 \\
\quad 90th Percentile (\%) & 39.86 & 88.59 & 100.00 & 100.00 \\
\quad Max (\%) & 84.44 & 100.00 & 100.00 & 100.00 \\
\hline
\multicolumn{5}{l}{\textbf{Average Performance}} \\
\quad Jobs Submitted Daily (thousand) & 2.53 & 5.06 & 7.59 & 10.12 \\
\quad Jobs Queued (\%) & 0.00 & 20.15 & 44.86 & 75.64 \\
\quad Job Queue Time (hours) & --- & 0.59 & 2.58 & 6.54 \\
\hline
\end{tabular}
\label{tab:whole-facility-colocation}
\end{table}

Some metrics aggregated over the whole year are presented in Table~\ref{tab:whole-facility-colocation}.
Although maximum utilization reached 100\% in most of the simulations -- i.e. all nodes were concurrently utilized -- power consumption only reached a maximum $\sim$7.30 MW, or approximately 73\% of the data center rated power design.
These results are strictly tied to the workload power profiles assumed in the simulation, and are consistent with the workloads in Sections~\ref{sec:exp_finetuning_LL} and~\ref{sec:exp_finetuning_SF}, where none of the jobs consistently consumed close to the nodes' TDP. 
These results however do raise questions on available resource headroom at the whole-facility level, and whether that should be considered when sizing infrastructure.
For instance, if it can be shown that IT infrastructure consistently consumes below its rated power, upstream equipment inside the data center or at the distribution-level might be undersized without compromising operation.
The bottom section in Table~\ref{tab:whole-facility-colocation} shows the average daily job count to reach the expected utilization, ranging from 2,530 jobs per day at 20\% utilization to 10,120 jobs per day at 80\%.
Not all jobs were however executed without queue.
While 20\% average utilization resulted in no queued jobs, and 40\% resulted in average queue time slightly over 30 minutes, 60 and 80\% average utilization resulted in significant jobs experiencing queues, with average queue time higher than 2 and 6 hours, respectively.

Another notable aggregated metric in Table~\ref{tab:whole-facility-colocation} is the peak-to-average ratio (PAR), defined as the ratio between the peak and average consumption over the whole simulation period, which is used to indicate variability of a load and its potential negative impact on the grid -- a PAR value equal to 1 is ideal.
In this case, PAR consistently decreased as utilization increased, which was the result of a higher utilization reducing the effect of hourly variations in user-behavior, submitted jobs and therefore load variability.

\begin{figure}[ht]
     \centering
     \begin{subfigure}[b]{0.5\textwidth}
         \centering
         \includegraphics[width=\textwidth]{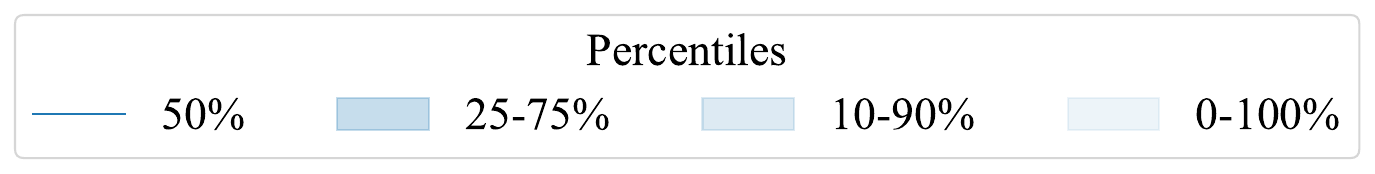}
     \end{subfigure}
     \hfill \\
     \begin{subfigure}[b]{0.245\textwidth}
         \centering
         \includegraphics[width=\textwidth]{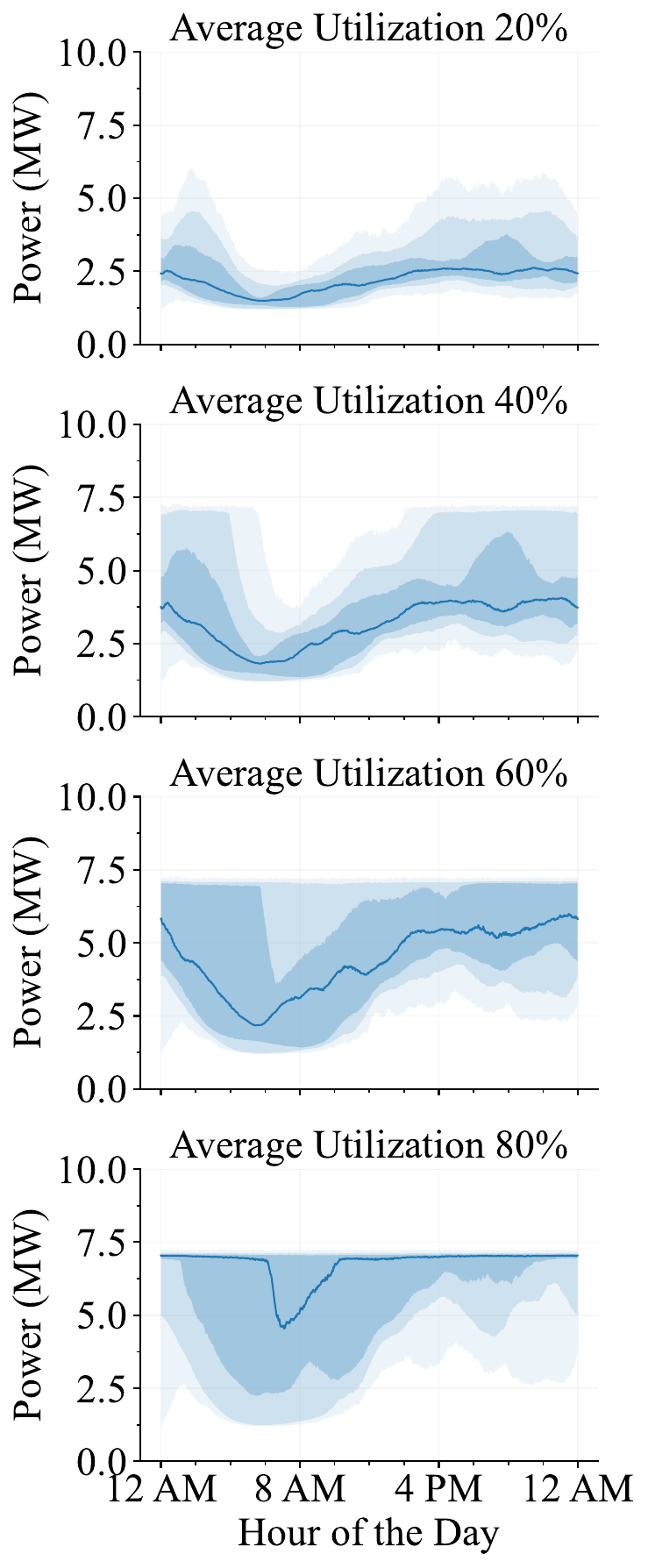}
         \caption{}
         \label{fig:colocation-day}
     \end{subfigure}
    \hfill
      \begin{subfigure}[b]{0.745\textwidth}
         \centering
         \includegraphics[width=\textwidth]{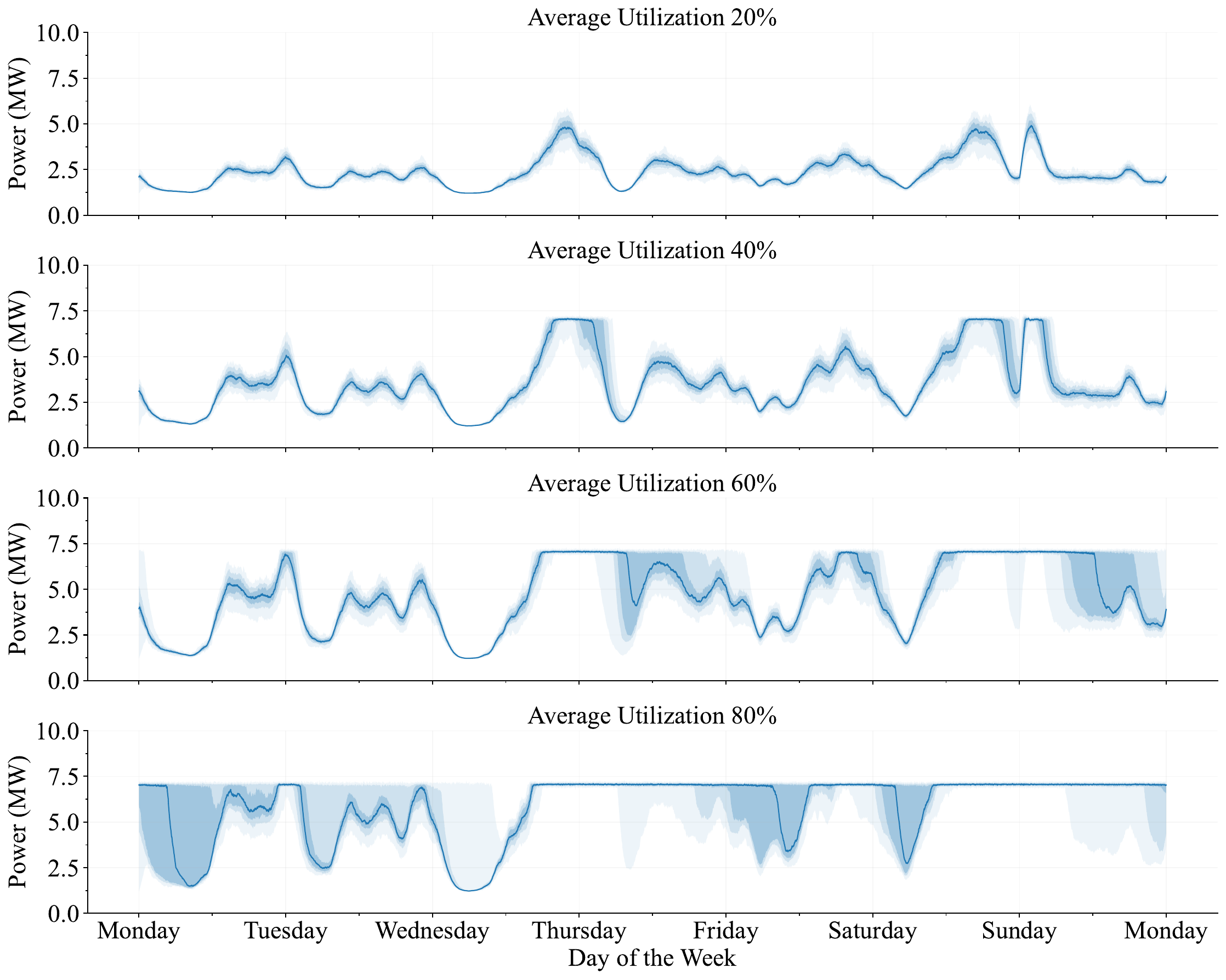}
         \caption{}
         \label{fig:colocation-week}
     \end{subfigure}
        \caption{Facility-level power consumption profile distributions for a 10 MW colocation data center at varying utilization levels.~\ref{fig:colocation-day}: by time of the day;~\ref{fig:colocation-week}: by day of the week. Distributions were generated from year-long simulations at 1-minute timesteps. The solid lines represent the median value, and the shaded areas different percentiles.}
        \label{fig:whole-facility-colocation-power}
\end{figure}

Figure~\ref{fig:whole-facility-colocation-power} shows distributions in the data center power consumption profiles, with each row representing a different average utilization assumption.
Figure~\ref{fig:colocation-day} shows the daily load distribution.
In general, job submission tended to increase after 8 AM, with peaks after 4 PM.
This diurnal pattern was more visible for moderate utilization levels -- 40 and 60\%.
At 20\% the load was dominated by the servers' idle consumption, whereas at 80\%, the load saturated due to sustained maximum utilization and job queuing.
This saturation in utilization and power consumption was also the reason why PAR decreased at higher average utilization levels in Table~\ref{tab:whole-facility-colocation}.
Conversely, Figure~\ref{fig:colocation-week} shows load distribution by time of the week.
We observed clear patterns, with highest consumption on Thursday and Saturday evening, bleeding into the following days which maintained higher utilization levels.
Since we assumed consistent facility utilization distributions with time of the week, the variability in Figure~\ref{fig:colocation-day} was due to the seasonal distribution based on NLR's HPC.

\subsection{Inference Data Center}\label{sec:wf_results_inference}

\begin{table}[t]
\caption{Inference data center aggregate metrics across the whole year for different target utilization levels.}
\small
\centering
\begin{tabular}{lcccc}
& \multicolumn{4}{c}{\textbf{Average Utilization}} \\
\hline
 & 20\% & 40\% & 60\% & 80\% \\
\hline
\multicolumn{5}{l}{\textbf{Power}} \\
\quad Mean (MW) & 0.26 & 0.39 & 0.53 & 0.66 \\
\quad Std Dev (MW) & 0.06 & 0.11 & 0.15 & 0.13 \\
\quad Median (MW) & 0.24 & 0.36 & 0.49 & 0.68 \\
\quad 90th Percentile (MW) & 0.35 & 0.58 & 0.80 & 0.80 \\
\quad Max (MW) & 0.46 & 0.80 & 0.80 & 0.80 \\
\quad Peak-to-Avg Ratio & 1.79 & 2.04 & 1.52 & 1.21 \\
\hline
\multicolumn{5}{l}{\textbf{Utilization}} \\
\quad Mean (\%) & 19.86 & 39.85 & 59.87 & 79.83 \\
\quad Std Dev (\%) & 8.38 & 16.97 & 22.78 & 19.68 \\
\quad Median (\%) & 17.54 & 35.44 & 54.74 & 82.81 \\
\quad 90th Percentile (\%) & 33.33 & 67.02 & 100.00 & 100.00 \\
\quad Max (\%) & 49.82 & 100.00 & 100.00 & 100.00 \\
\hline
\multicolumn{5}{l}{\textbf{Average Performance}} \\
\quad Daily Prompts (billion) & 0.30 & 0.60 & 0.93 & 1.42 \\
\quad Incoming Request Rate (thousand prompts/second) & 3.43 & 6.95 & 10.78 & 16.39 \\
\quad Effective Request Rate (thousand prompts/second) & 3.43 & 6.95 & 10.49 & 14.01 \\
\quad Incomplete Request Rate (thousand prompts/second) & 0.00 & 0.00 & 0.30 & 2.38 \\
\hline
\end{tabular}
\label{tab:whole-facility-inference}
\end{table}

We simulated operation at a 1 MW data center serving LLM inference in production (online) mode.
In accordance with the benchmark tests in Section~\ref{sec:exp_online_inference} and the power samples in Figure~\ref{fig:online_inference_series_rate}, we assumed Llama-3 70B model instances to serve user requests answering coding and conversation questions, respectively.
To inform the data center utilization, we used distributions published by Microsoft Azure~\cite{azure_llm_dataset, azure_llm_paper}.
These distributions include a probability function dictating the fraction of coding and conversation prompts -- 38.1 and 61.9\%, respectively -- as well as functions to shape how the request rate (prompts/second) varies with time -- hourly and by day of the week. Similarly to the colocation use case, since the utilization data does not span a whole year, NLR own data center's job log was used to inform the seasonality in facility utilization.

As explained in Section~\ref{sec:model_structure}, DIPLOEE also requires parameters to establish how prompts are distributed among model instances.
In this example, we made the assumption that end-to-end response latency -- $\lambda _{i, \max}$ in Equation~\ref{eq:l_max} -- was to be maintained below a certain threshold to favor a positive user experience.
Specifically, based on Figure~\ref{fig:online_inference_performance}, we set the maximum request rate -- $\dot{R}_{i, \max}$ in Equation~\ref{eq:l_max} -- for coding and conversation prompts to 100 and 50 prompts per second respectively, limiting latency below 10 seconds.
We expect this choice in $\dot{R}_{i, \max}$ to significantly affect the volume of requests that can be completed by the data centers, the server utilization, and the resulting power profile.
We plan to directly evaluate this tradeoff as part of future work.

\begin{figure}[t]
     \centering
     \begin{subfigure}[b]{0.5\textwidth}
         \centering
         \includegraphics[width=\textwidth]{figures/colocation-legend.pdf}
     \end{subfigure}
     \hfill \\
     \begin{subfigure}[b]{0.245\textwidth}
         \centering
         \includegraphics[width=\textwidth]{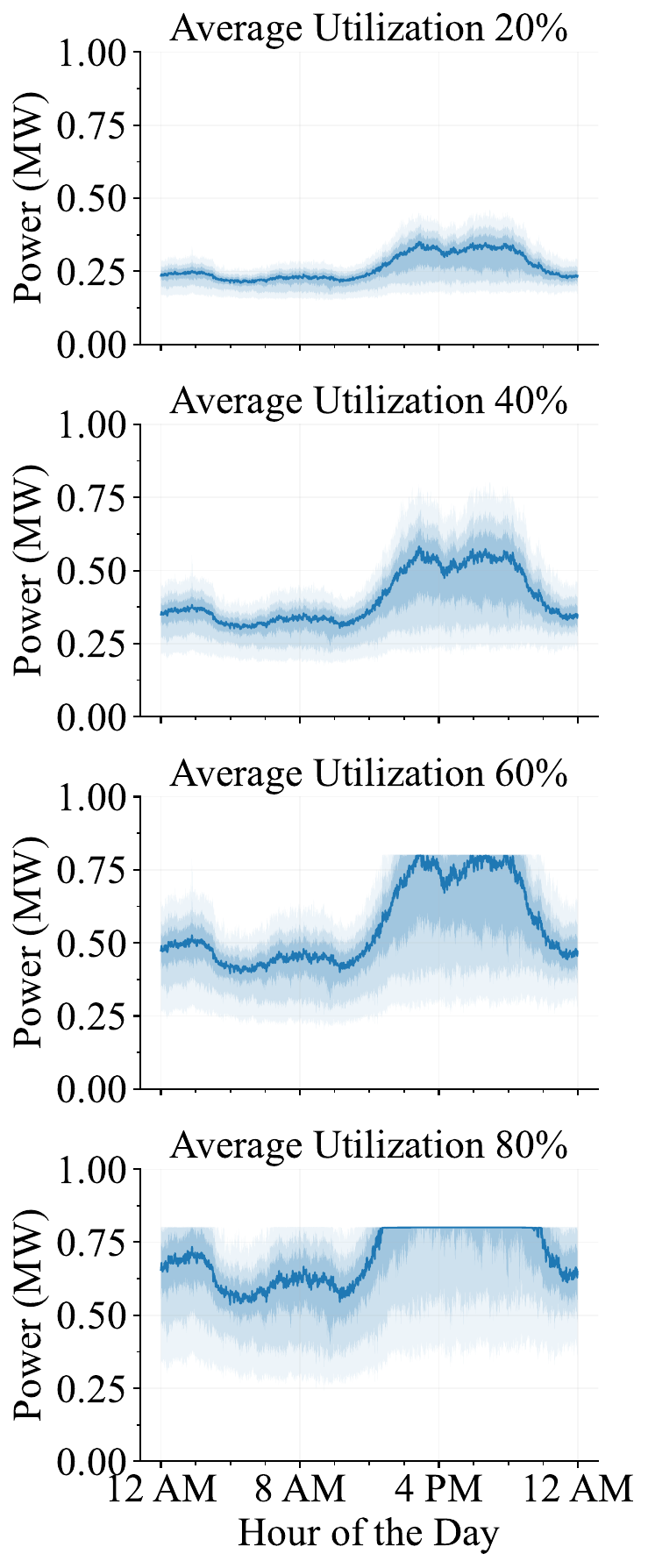}
         \caption{}
         \label{fig:inference-day}
     \end{subfigure}
    \hfill
      \begin{subfigure}[b]{0.745\textwidth}
         \centering
         \includegraphics[width=\textwidth]{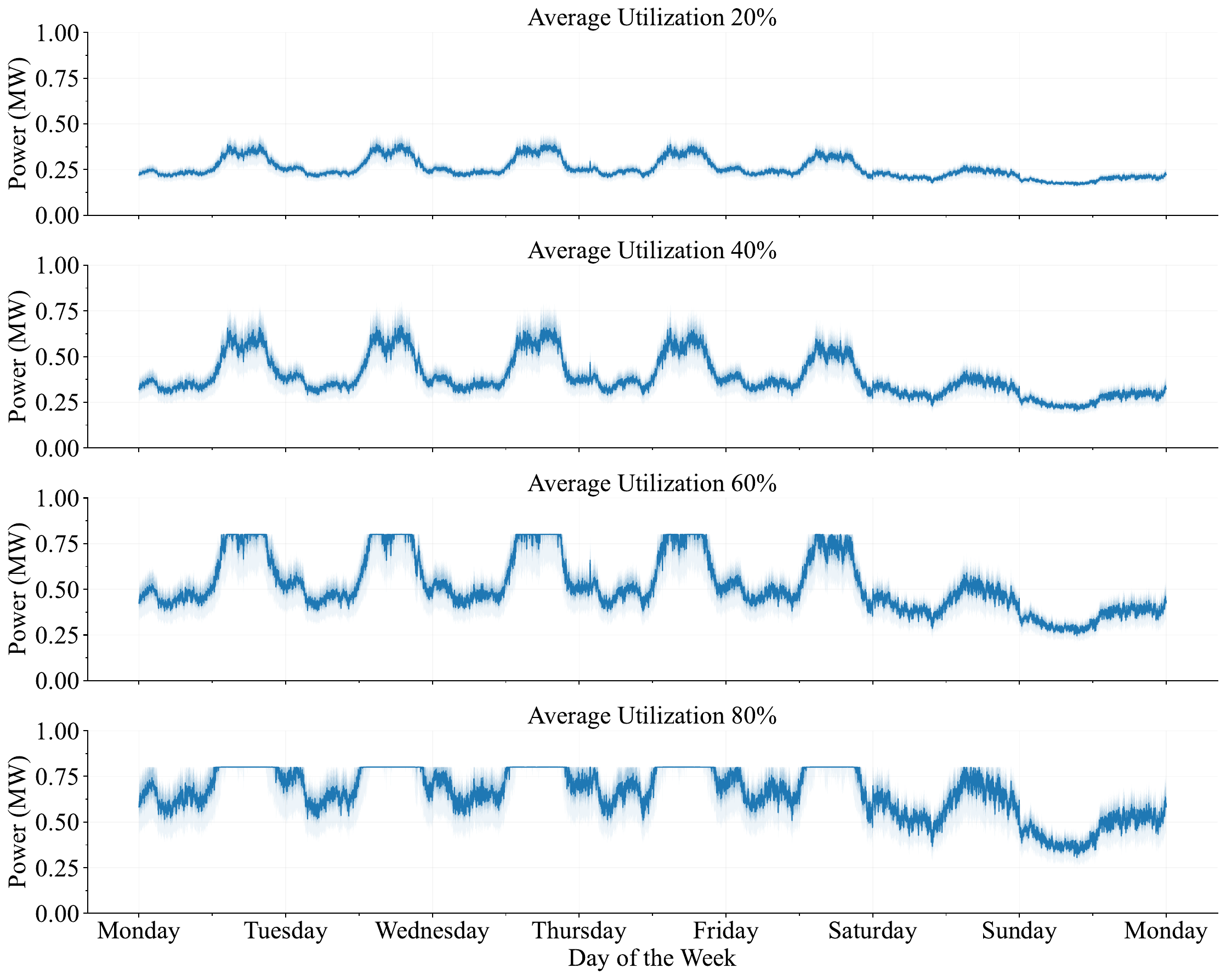}
         \caption{}
         \label{fig:inference-week}
     \end{subfigure}
        \caption{Facility-level power consumption profile distributions for a 1 MW inference data center at varying utilization levels.~\ref{fig:inference-day}: by time of the day;~\ref{fig:inference-week}: by day of the week. Distributions were generated from year-long simulations at 1-minute timesteps.  The solid lines represent the median value, and the shaded areas different percentiles.}
        \label{fig:whole-facility-inference-power}
\end{figure}

Similarly to Section~\ref{sec:wf_results_colocation}, we assumed the same node architecture as NLR's HPC.
Excluding cooling and auxiliary nodes, such an architecture could support a total of 284 nodes in a 1 MW data center.
In accordance with our benchmark tests, each model instance was served on one node. 
The nodes were allocated to host either coding or conversation model instances based on their respective probability, as defined in Equation~\ref{eq:inference_nodes_allocation}, resulting in 218 nodes allocated to conversation requests and 67 for coding.

Aggregated result metrics are presented in Table~\ref{tab:whole-facility-inference}.
Similarly to colocation, although maximum utilization reached 100\% in all utilization scenarios besides 20\%, the maximum power only reached 80\% -- 7 percentage points higher than the colocation case -- of the rated 1-MW power, due to the consumption on individual servers rarely reaching its maximum rate.
In the bottom section of the table, the average daily prompts added to 300 million at 20\% average utilization, corresponding to approximately 3.43 thousand prompts per second and requiring 57 model instances; at 80\% utilization, 1.42 billion prompts were received each day -- on average 16.39 thousand prompts per second. 
Due to the limitation on the number of nodes and latency constraints, approximately 3\% of requests were not processed at 60\% average utilization, 15\% at 80\%.

Finally, peak-to-average ratio (PAR) values showed an interesting trend: the 40\% utilization scenario showed a worse (higher) PAR value than 20\%, indicating more variability, while at higher utilization PAR decreased due to sustained resources saturation.
Compared to colocation, PAR values for the inference use case were 9\% lower on average, potentially suggesting that inference facilities could result in less grid disruption.
However, the impact of resources saturation is arguably more detrimental to inference data centers, which provide real-time and often critical services, than to colocation data centers, where users expect their job to spend some time in the queue.
Therefore, while inference data centers might pose less stability risks to the grid than colocation at similar utilization levels, they are more likely to operate at lower average utilization to avoid a degraded level of service.

Figure~\ref{fig:whole-facility-inference-power} shows distributions in the data center power consumption profiles.
Figure~\ref{fig:inference-day} shows the daily load distribution, which clearly followed a diurnal pattern, with requests ramping up after 9 AM and decreasing after 10 PM.
Similarly to  the colocation use case, this diurnal pattern became less prevalent with higher utilization, because the servers' utilization saturated.
Figure~\ref{fig:inference-day} shows load distribution by time of the week.
These distributions also showed alignment with the typical "business week", with highest load between Monday and Friday, and lower consumption during the weekend.
Seasonal variations were also visible through the different shadings.


\section{Conclusion}
\label{sec:conclusion}

This work presented a bottom-up framework connecting high-resolution, node-level power measurements of representative GenAI workloads to whole-facility data center energy modeling for infrastructure planning and operational studies.
By combining reproducible benchmarks on commercial-scale GPU systems with a discrete-event simulation model including job submission, scheduling, and execution profiles, we bridged a critical gap between fine-grained workload characterization and facility-level demand profiles that are meaningful for generation, transmission, and distribution planning, as well as for data center design and operation.

Based on the presented node-level AI workload measurements, we demonstrated that both training and inference jobs exhibit pronounced power transients that can vary widely according to the type of model and the number of nodes used for computation.
For LLM and image-generation training workloads, power scaled linearly with node count (under the tested configurations), while runtime was strongly influenced by the increase in global batch size at higher node counts.
For online inference workloads, server-level power seemed to saturate at relatively modest request rates, highlighting that performance might impact power in a diminishing trend beyond certain operating points, and underscoring the importance of further investigating the tradeoffs between customer experience and energy consumption.

In contrast, the facility-level simulation results underscored the correlation between data center utilization -- and customer-behavior -- and whole-facility consumption, with clear diurnal, weekly, and seasonal patterns.
The variability in whole-facility load profiles due to these patterns was clearly visible at low to moderate average utilization levels (20, 40, 60\%).
At higher utilization (80\%), the load often saturated, resulting in profiles that were more grid-stable at the cost of a deteriorated customer experience, particularly for inference data centers.
Interestingly, even when all servers were being utilized, the facility power consumption only reached 73 and 80\% of the data center rated design, indicating an opportunity for optimizing the cost of data center auxiliary infrastructure.


The proposed framework is subject to a few limitations. 
First, measurements were conducted on a specific hardware setup and using a limited set of algorithm configurations and workload types.
Different accelerators, model architectures, or even interconnect technologies may exhibit distinct power-performance characteristics.
Second, the facility simulations did not represent auxiliary loads (cooling, power distribution losses, and other non-IT loads), although these might constitute less than 10\% of the total facility load for large, state-of-the-art AI data centers. 
Third, utilization and arrival-rate distributions adapted from external datasets may not fully reflect the operational diversity of commercial hyperscale or enterprise environments.
Future work will address these limitations by expanding the public dataset to include additional models, tasks, number of model parameters, algorithm hyperparameters, and possibly hardware platforms.
The facility-level model will be expanded by integrating thermal and electrical infrastructure models to capture cooling and power delivery dynamics, as well as smart strategies for job scheduling.

By providing an open-source dataset of node-level power consumption profiles and a clear methodology for scaling up to the facility level, we aim to provide a foundation for data-driven planning of the rapidly expanding AI compute infrastructure, and assist decision makers in the data center space with infrastructure planning and load forecasting.


\clearpage
\newpage
\appendix

\section{GPU and CPU Stress and Idle Tests}
\label{appA}

\begin{figure}[ht]
    \centering
    \includegraphics[width=0.48\linewidth]{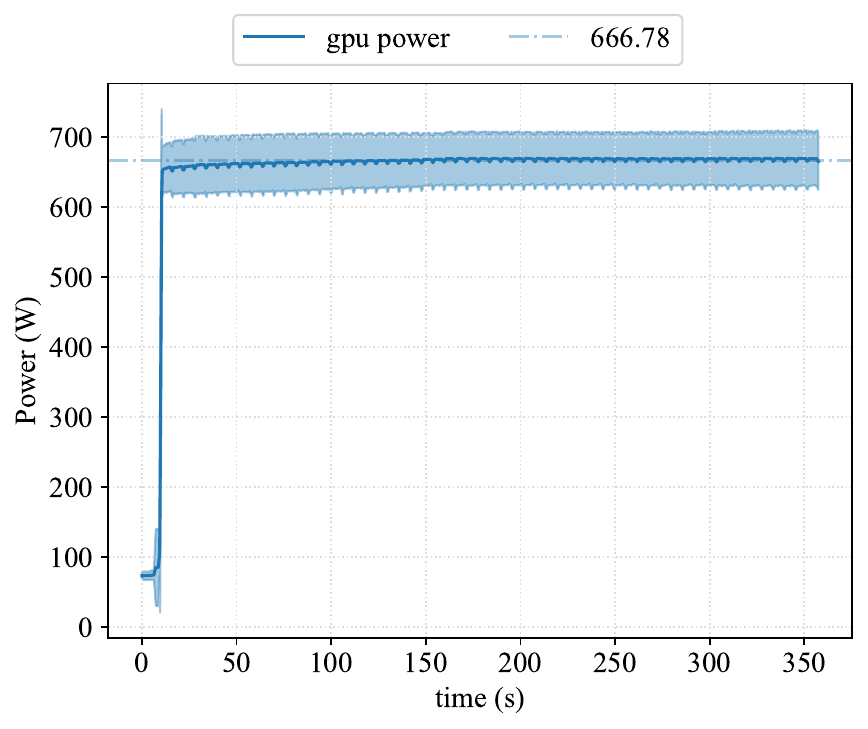}
    \hfill
    \includegraphics[width=0.48\linewidth]{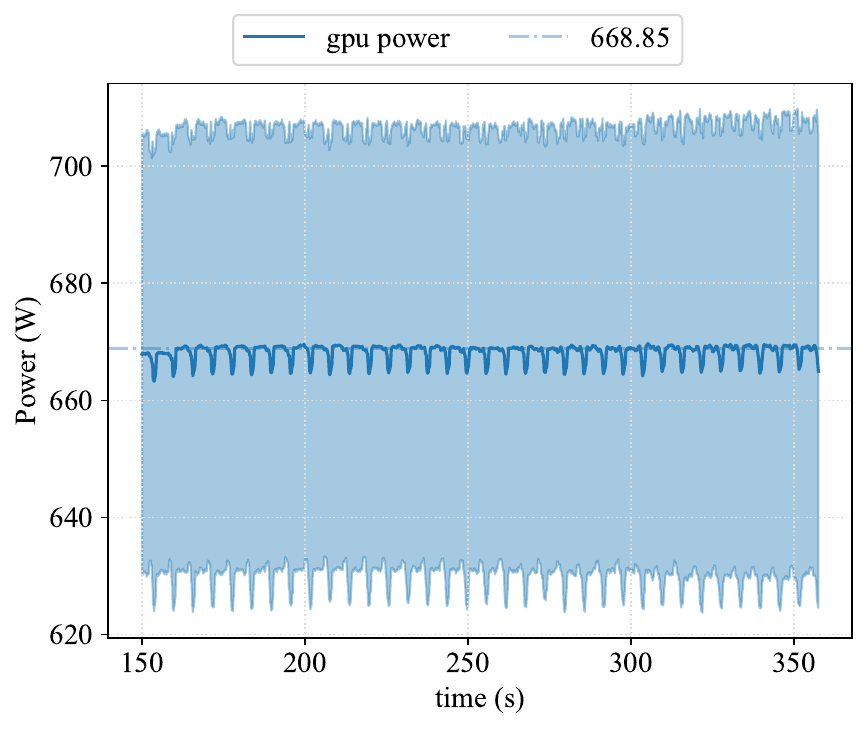}
    \caption{GPU power when stressed using \texttt{gpu-burn}~\cite{gpu-burn}, median value, and 99\% confidence bound (z score: 2.58). Mean value of 668.2 W ($\pm$ 1.4 W std) after warm up of 150 seconds.}
    \label{fig:gpu-stress}
\end{figure}

\begin{figure}[ht]
    \centering
    \includegraphics[width=0.5\linewidth]{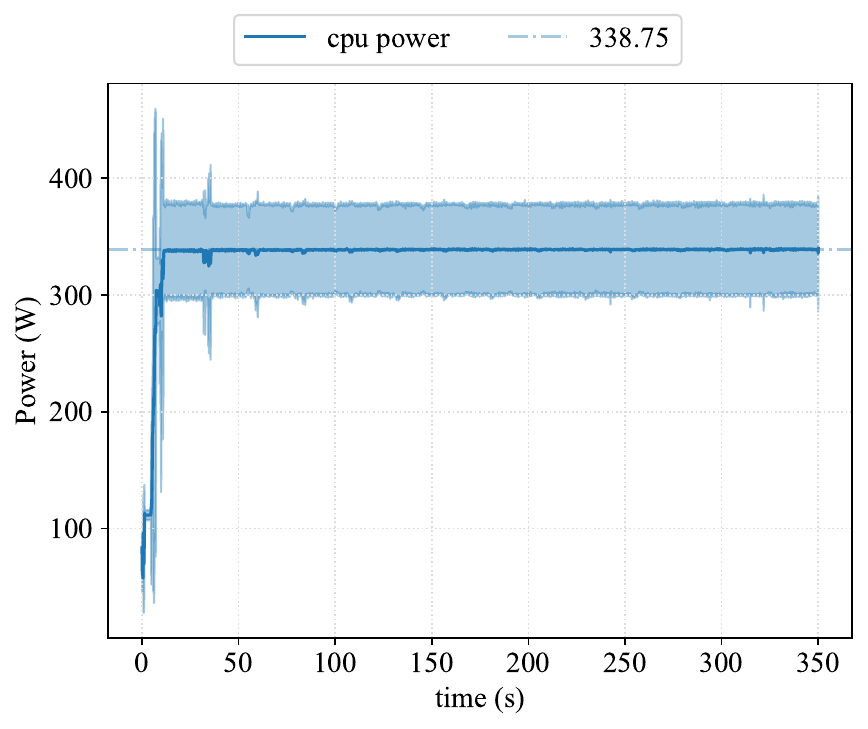}
    \caption{CPU power for a dense matrix multiplication kernel, median value, and 99\% confidence bound (z score: 2.58). Mean value of 338.6 W ($\pm$ 1.0 W std) after warm up of 11 seconds.}
    \label{fig:cpu-stress}
\end{figure}

To evaluate the power and thermal performance bounds of the computing system, 
we conducted dedicated stress and idle tests for both the GPUs and CPUs. 
The test corresponds to a full-load stress scenario designed to drive the hardware to its maximum sustained utilization, allowing for measurement of peak power and thermal usage under high-demand conditions. 
For the GPU tests, the NVIDIA H100 accelerators were subjected to continuous full-precision matrix and tensor operations, representative of deep learning workloads. 
Similarly, the CPU stress tests involved all-core utilization of the AMD EPYC 9554 processors using a sequence of dense matrix multiplications aiming to saturate available threads and cache resources. 
In contrast, the idle tests captured the minimum baseline power draw behavior when the system is not executing tasks.

\begin{figure}[ht]
    \includegraphics[width=0.48\linewidth]{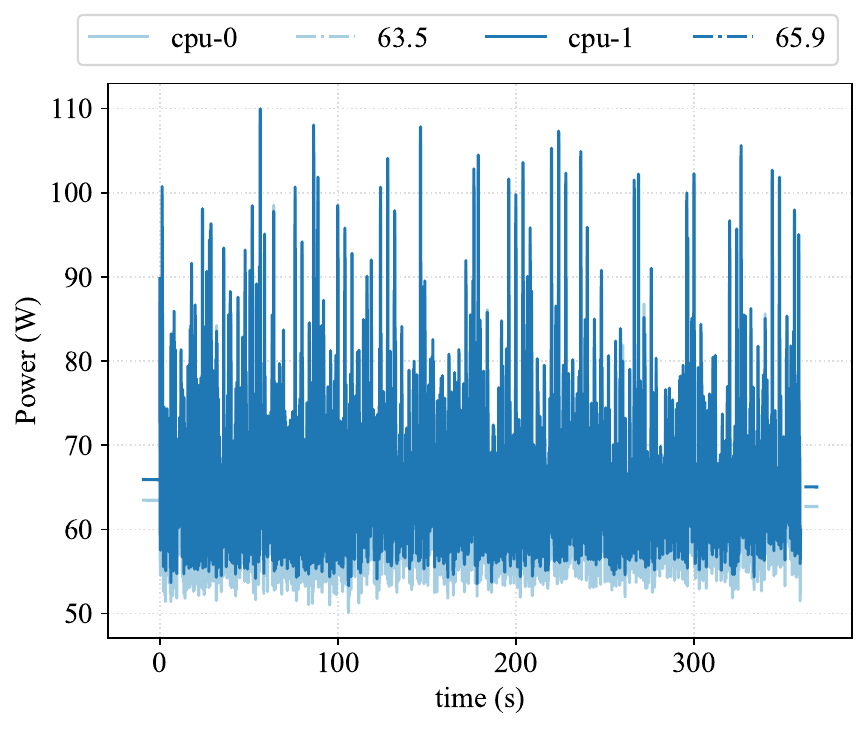}
    \centering
    \includegraphics[width=0.48\linewidth]{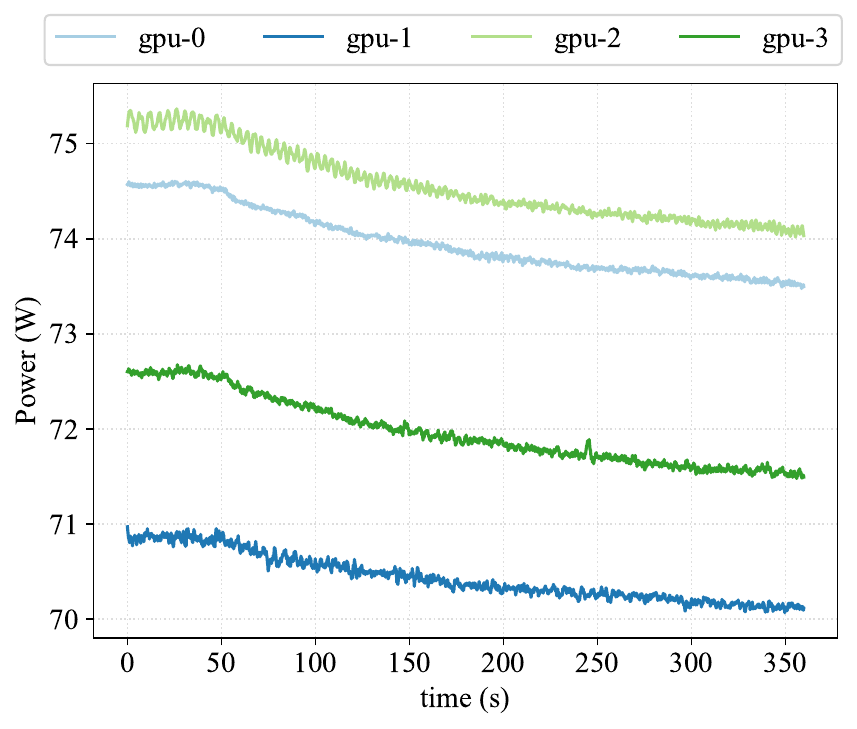}
    \caption{CPU (left) and GPU (right) idle power on a Kestrel GPU node. We show the linear regression for the CPU data.}
    \label{fig:gpu-idle}
\end{figure}

Figure~\ref{fig:gpu-stress} shows the stress test results using \texttt{gpu-burn}~\cite{gpu-burn}. The average value was about 7\% lower than the TDP from the manufacturer~\cite{nvidiah100}.
Figure~\ref{fig:cpu-stress} shows the results of the sequence of dense matrix multiplications. 
All matrices are 8192-by-8192 with entries uniformly random distributed in the $[0,1)$ interval which successfully put power levels inside the 320--400W configurable TDP range as reported by the manufacturer~\cite{amdepicgenoa}.
Figure~\ref{fig:gpu-idle} shows idle power profiles for both GPUs and CPUs in a GPU node of Kestrel. 
The two CPU sockets operated at slightly different mean power, and the linear regression shows the profile seemed to be constant in time. 
The power draw associated to each GPU in a node, however, oscillated in phase, each curve following a different mean. When comparing several GPU nodes on Kestrel, we noticed that: (1) each idle CPU socket spent 64.1 W on average, with a standard deviation of $\pm$ 4.8 W; (2) one of the sockets reported higher (1-3 W) power usage than the other inside the same node; (3) each idle GPU spent 72.5 W on average, with a standard deviation of $\pm$ 0.1 W; (4) each GPU in a node operated at a slightly distinct average power level, with a few units of Watts difference, with power curves showing in-phase oscillations.

\begin{figure}[ht]
\centering
    \includegraphics[width=0.5\linewidth]{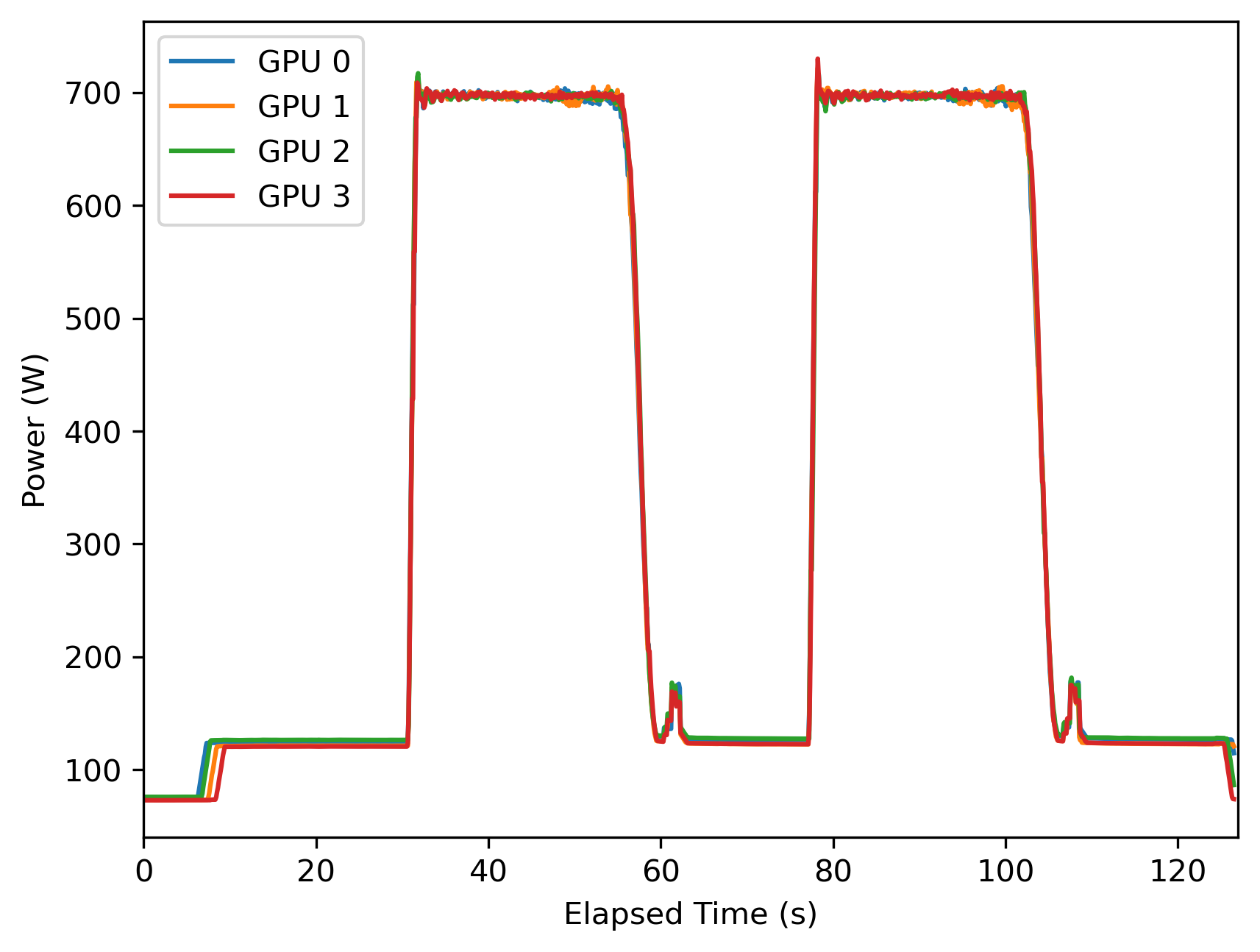}
    \caption{Power consumption of a single node, four device HPL-NVIDIA run with two matrix solves.}
    \label{fig:HPL-GPU}
\end{figure}

In addition to the dedicated CPU and GPU burn tests, we report here the power consumption of a GPU-enabled High Performance Linpack (HPL) microbenchmark on NLR's H100 GPU nodes via HPL-NVIDIA 24.03.0. HPL is a dense linear algebra-based benchmark that measures the floating-point operations per second (FLOPS) achieved while solving a system of linear equations via LU-factorization. This test serves as an approximate baseline for user-achievable maximum power consumption in the compute-bound regime on Kestrel's H100 devices. Figure~\ref{fig:HPL-GPU} shows the power profile during the HPL-NVIDIA test for two matrix solves (corresponding to the two ~695 W regions) on one node with four H100 GPU devices. These two solves achieved 168.6 TFLOPs and 169.5 TFLOPs respectively, with 42.15 TFLOPS and 42.38 TFLOPS per GPU device. These results are consistent with previously reported HPL-NVIDIA results on H100 GPUs.~\cite{BenchmarkingAITennesseeInitiative}


\clearpage

\section*{Acknowledgment}



This work was authored by the National Laboratory of the Rockies for the U.S. Department of Energy (DOE). Funding provided by the Advanced Scientific Computing Research (ASCR) program. The views expressed herein do not necessarily represent the views of the DOE or the U.S. Government.
This research was performed using computational resources sponsored by the U.S. Department of Energy's Office of Critical Minerals and Energy Innovation and located at the National Laboratory of the Rockies.


\sloppy
\bibliographystyle{ieeetr}
\bibliography{references}

\end{document}